\documentclass[11pt]{article}
\usepackage{graphics}
\usepackage{epsfig}

\newcommand{\be}{\begin{eqnarray}}
\newcommand{\ee}{\end{eqnarray}}

\def\ll#1{\left#1}
\def\r#1{\right#1}
\def\fr{\frac{1}{2}}

\def\mref#1{(\ref{#1})}

\def\p{\partial}

\def\bd{\begin{displaymath}}
\def\ed{\end{displaymath}}

\def\ba#1{\begin{array}{#1}}
\def\ea{\end{array}}
\def\nn{\nonumber}
\newfont{\Bbb}{msbm10 scaled 1200}

\begin{document}

\pagestyle{empty}

\begin{center}

{\LARGE\bf Semiclassical wave functions in billiards built on classical trajectories. Energy quantization,
scars and periodic orbits\\[0.5cm]}

\vskip 12pt

{\large {\bf Stefan Giller{$\dag$} and Jaros{\l}aw Janiak{$\ddag$}}}

\vskip 3pt

{$\dag$}Jan D{\l}ugosz University in Czestochowa\\
Institute of Physics\\
Armii Krajowej 13/15, 42-200 Czestochowa, Poland\\
e-mail: stefan.giller@ajd.czest.pl\\
$\ddag$Theoretical Physics Department II\\
University of {\L}\'od\'z,\\
Pomorska 149/153, 90-236 {\L}\'od\'z, Poland\\
e-mail: j.janiak@yahoo.pl

\end{center}
\vspace{6pt}

\begin{abstract}
A way of construction of semiclassical wave function (SWF) based on the Maslov - Fedoriuk approach
is proposed which appears to be appropriate also for systems with chaotic classical limits. Some classical constructions
called skeletons are considered. The skeletons are generalizations of Arnolds' tori able to gather chaotic dynamics.
SWF's are continued by caustic singularities in
the configuration space rather then in the phase space using complex time method. The skeleton formulation provides us
with a new algorithm for the semiclassical approximation method which is applied to
construct SWF's as well as to calculate energy spectra for the circular and rectangular billiards as well as to construct
the simplest SWF's and the respective spectrum for the Bunimovich stadium. The scar phenomena are considered and a possibility of their description by the skeleton
method is discussed.

\end{abstract}

\vskip 3pt
\begin{tabular}{l}
{\small PACS number(s): 03.65.-w, 03.65.Sq, 02.30.Jr, 02.30.Lt, 02.30.Mv} \\[1mm]
{\small trajectories, chaotic dynamics, quantum chaos, scars}
\end{tabular}

\newpage

\pagestyle{plain}

\setcounter{page}{1}

\section{Introduction}
\hskip+2em The semiclassical approximation is widely used and by this it is the well known method of approximation in quantum
physics. There are two basic formulations of the method - the one based on the wave function formulation of quantum
mechanics \cite{1,4} and the other - on the Feynman paths integral \cite{6,2}. While both the formulations of quantum mechanics are
known to be equivalent the general wave function formulation of the semiclassical approximation \cite{4} is considered
to be not applicable in higher dimensional quantum problems which classical limits are chaotic.

In fact it is a common convince that the only way to formulate the
method in the last cases is the Gutzwiller approach based on the Feynman paths integral \cite{3}. This convince follows
from also
a common believe that the wave function formalism can be applied only in the cases when the classical limit of the
quantum problem is the integrable one i.e. if the classical motion is set on the Arnold tori \cite{5,7} on which
the semiclassical wave functions (SWF) are constructed. As the main
argument for such a believe the KAM theorem \cite{5,7} is invoked which claims that the Arnold tori structure of the
classical phase space disappears if classical systems become non-integrable. It is argued that because of that the only
phase space finite structure which
still survive in chaotic motion of the classical system are periodic orbits which therefore provide us with a
skeleton on which the Gutzwiller formula is built.

However one can criticise the point of view that the existence of the Arnold tori structure of the phase space is
necessary for a possibility to construct the semiclassical wave function, i.e. that Arnold's tori provide us
with a unique support for such a construction. Such a definite conclusion does certainly not follow from the original
local approach of Maslov and Fedoriuk to construct SWF's \cite{4}.

On the other hand while the results provided by the Gutzwiller method are very rich and appreciated, particularly when
energy spectra of the chaotic systems are considered (see for example \cite{18}), the method itself does not allow us for constructing and discussing
properties of the wave functions involved in the problems considered. In fact the wave functions are found in such
cases by different methods mostly numerically. It is just due to such numerical calculations of the wave functions
\cite{9,11}
that a phenomenon of scars has been discovered \cite{10} which existence in the wave function patterns is still waiting
for its full explanation \cite{12,20,25} (see also other papers cited in \cite{12}).

Billiards while a non-analytic motion area are well known however as examples of the non-integrable two dimensional
systems except the known cases of the elliptical and rectangular billiards. They are widely considered as a simple field
of experimental \cite{13,14} as well as theoretical \cite{15,16,17,26} (and papers cited there) and computational
investigations \cite{9,24,22,23} allowing to apply many different methods (see Sarnak's lecture \cite{21} and \cite{27}
of the same author for an extensive review of the respective theoretical methods covering also billiards manifolds).

In this paper we are going to
develop the SWF formalism which can be applied at least in principle to non-integrable cases of the two dimensional
motions in billiards and which can be easily extended to higher dimensions.

Essentially our approach is initially very close to the one of Maslov and Fedoriuk \cite{4}. The main difference
between them is
in a treatment of crossing the singular points of the SWF's set on caustics. Namely, instead of making the canonical
phase
space variable transformations accompanied by the Fourier transformations of the SWF's to move through the caustic
points we apply the analytical continuation on the complex time plane to both the SWF's and the classical trajectories.
This greatly simplifies the corresponding procedure in comparison with the Maslov and Fedoriuk treatment. It is
the exceptional role played by the time variable in the semiclassical limit of the Schr\"odinger equation which permits
us for such simplification. Because of this the SWF can be considered as depending effectively on the time variable only
while the remaining variables plays the role of spectators.

Therefore the SWF's are first constructed locally on so called bundles of rays to satisfy vanishing boundary conditions.
Next they are matching to get a global semiclassical solution. This is done however with the help of an earlier constructed
set of reversible in time closed connected ray bundles called bundle skeleton which play a role of Arnold's tori except that a
number of ray bundles in the skeleton can be infinite.

This is just the notion of the ray bundles which allows us to catch a possible chaotic motion in the billiards
not to resign from considerations of a set of trajectories on which SWF's can be defined while the bundle skeleton
idea allows us to close the matching procedure of SWF's constructed locally.

The paper is organized as follows.

In the next section the Maslov - Fedoriuk method of the semiclassical wave function construction is reminded and
discussed.

In sec.3 the semiclassical wave function is considered as the classical objects which time evolution is described by
the classical equations of motion.

In sec.4 the construction of global SWF's in billiards is given.

In sec.5 the circular billiards is considered to demonstrate how the method works in the case of the presence of caustic.

In sec.6 the rectangular billiards is considered as the case deprived of a caustic.

In sec.7 our method is applied to the Bunimovich stadium to show its usefulness in describing
the so called bounces ball modes.

In sec.8 we discuss a possibility to describe by the skeleton idea the scar phenomenon considering such a scar formed
around the horizontal periodic orbit in the Bunimovich stadium.

In sec.9 the results of the paper are summarized and some limitations of the method are discussed.

There are four appendixes attached to the paper which justify the main assumptions used in the construction of the global
SWF's on skeletons.

\section{Semiclassical wave function expansion for $n$-D stationary Schr\"odinger equation}

\hskip+2em Consider the $n$-dimensional stationary Schr\"odinger equation:
\be
\bigtriangleup\Psi({\bf r})+\lambda^2\frac{2m}{\hbar^2}(E-V({\bf r}))\Psi({\bf r})=0
\label{1}
\ee
with a potential $V({\bf r}),\;{\bf r}\in R_n$ confining a point particle with a mass $m$ and containing a formal dimensionless parameter
$\lambda>0$. For a convenience we shall put further $\hbar=1$ and $m=1$. The Schr\"odinger equation is recovered by
putting $\lambda=1$ in \mref{1}.

We would like to construct a solution to Eq.\mref{1} using the idea of Maslov {\it et al} \cite{4} and considering the wave function $\Psi({\bf r})$
as defined on families of classical trajectories a dynamic of which is given by the classical Hamiltonian $H=\fr{\bf p}^2+V({\bf r})$ and which
carry an energy $E_0$ all.

Such families are constructed locally in the following way. In $R_n$ we choose a $n-1$-D hypersurface $\Sigma_{n-1}$ on which the initial momenta
${\bf p}({\bf r}_0),\;{\bf r}_0\in\Sigma_{n-1}$, will be defined so that the pair
$({\bf r}_0,{\bf p}({\bf r}_0)),\;{\bf r}_0\in\Sigma_{n-1}$ will serve as initial data for the trajectory
${\bf r}(t)={\bf f}({\bf r}_0,{\bf p}({\bf r}_0);t)$. Then the transformation: ${\bf r}\to(t,{\bf r}_0)\to(t,s_1,...,s_{n-1}),
\;{\bf r}\equiv(x_1,...,x_n),\;{\bf r}_0\equiv(x_{0,1}(s_1,...,s_{n-1}),...,\linebreak x_{0,n}(s_1,...,s_{n-1}))$,
(($s_1,...,s_{n-1}$) parameterize the hypersurface $\Sigma_{n-1}$) is one-to-one up to a caustic surface $C_{n-1}$ on which the
Jacobean (${\bf f}({\bf r}_0,{\bf p}({\bf r}_0);t)\equiv{\bar{\bf f}}(t,s_1,...,s_{n-1})$):
\be
J(t,s_1,...,s_{n-1})=\ll|\frac{\p {\bar f}_i}{\p t},\frac{\p {\bar f}_i}{\p s_j}\r|
\label{4}
\ee
vanishes.

A $n$-dimensional domain $\Lambda_n$ of $2n$-dimensional phase space $R_{2n}$ made in this way by the hypersurface
$\Sigma_{n-1}$ and trajectories emerging from it is known as the Lagrange manifold \cite{5}.

Therefore in the variables $t,s_1,...,s_{n-1}$ the new wave function $\chi(t,s_1,...,s_{n-1})$ satisfies the following relation with the previous one:
\be
|\chi(t,s_1,...,s_{n-1})|^2=|\Psi({\bar{\bf f}}(t,s_1,...,s_{n-1})|^2|J(t,s_1,...,s_{n-1})|
\label{4a}
\ee

The particle momentum ${\bf p}$ on the trajectories ${\bf r}(t)={\bar{\bf f}}(t,s_1,...,s_{n-1})$ satisfies of course the equation:
\be
\frac{\p{\bar{\bf f}}(t,s_1,...,s_{n-1})}{\p t}={\bf p}({\bar{\bf f}}(t,s_1,...,s_{n-1}))
\label{5}
\ee
defining also the Jacobean evolution. Namely:
\be
\frac{\p}{\p t}\frac{\p{\bar{f}}_i(t,s_1,...,s_{n-1})}{\p t}=\sum_{j=1}^n
\frac{\p p_i}{\p x_j}\frac{\p{\bar{f}}_j(t,s_1,...,s_{n-1})}{\p t}\nn\\
\frac{\p}{\p t}\frac{\p{\bar{f}}_i(t,s_1,...,s_{n-1})}{\p s_l}=\sum_{j=1}^n
\frac{\p p_i}{\p x_j}\frac{\p{\bar{f}}_j(t,s_1,...,s_{n-1})}{\p s_l}\nn\\l=1,...,n-1
\label{6}
\ee

so that
\be
\frac{\p J(t,s_1,...,s_{n-1})}{\p t}=J(t,s_1,...,s_{n-1})\nabla {\bf p}({\bar{\bf f}}(t,s_1,...,s_{n-1}))
\label{7}
\ee

The above equation is just the Liouville theorem with the solution:
\be
J(t,s_1,...,s_{n-1})=J(s_1,...,s_{n-1})e^{\int_0^t\nabla {\bf p}({\bar{\bf f}}(t',s_1,...,s_{n-1}))dt'}
\label{7a}
\ee
where $J(s_1,...,s_{n-1})$ is the value of the Jacobean on the hypersurface $\Sigma_{n-1}$.

It is well known from the classical Hamiltonian mechanics \cite{5} that the action integral:
\be
S({\bf r},{\bf r}_0)=\int_{{\bf r}_0}^{\bf r}{\bf p}({\bf r}')d{\bf r}'
\label{7b}
\ee
taken on the Lagrange manifold $\Lambda_n$ is a point function of ${\bf r}$ and ${\bf r}_0$. Therefore taking ${\bf r}_0$ as a definite fixed
point of the hypersurface $\Sigma_{n-1}$ and denoting by $S({\bf r})$ the action function corresponding to this case we can complete
a definition of the wave function $\chi(t,s_1,...,s_{n-1})$ by the following equation:
\be
\Psi^\sigma({\bar{\bf f}}(t,s_1,...,s_{n-1}))=J^{-\fr}(t,s_1,...,s_{n-1}))
e^{\sigma\lambda iS({\bar{\bf f}}(t,s_1,...,s_{n-1})}\chi^\sigma(t,s_1,...,s_{n-1})\nn\\
\label{2}
\ee
where $\sigma=\pm$ is a signature of $\Psi^\sigma({\bf r})$.

Therefore the quantities involved in the above definitions satisfy the following equations:
\be
{\bf p}({\bf r})=\nabla S({\bf r})\nn\\
\fr{\bf p}^2({\bf r})+V({\bf r})-E_0=0\nn\\
\triangle(J^{-\fr}\chi^\sigma({\bf r}))+\sigma 2i\lambda J^{-\fr}({\bf r})
\nabla \chi^\sigma({\bf r})\cdot{\bf p}({\bf r})+2\lambda^2(E-E_0)J^{-\fr}({\bf r})\chi^\sigma({\bf r})=0\nn\\
{\bf r}={\bar{\bf f}}(t,s_1,...,s_{n-1})
\label{3}
\ee

By the variables $t,s_1,...,s_{n-1}$ the third of the last equations can be rewritten in the following form:
\be
\sigma 2i\lambda\frac{\p\chi^\sigma(t,s_1,...,s_{n-1},\lambda)}{\p t}+\nn\\
J^{\fr}\triangle\ll(J^{-\fr}\chi^\sigma(t,s_1,...,s_{n-1},\lambda)\r)+
\lambda^2(E-E_0)\chi^\sigma(t,s_1,...,s_{n-1},\lambda)=0
\label{11}
\ee
where a dependence of $\chi^\sigma(t,s_1,...,s_{n-1},\lambda)$ on $\lambda$ was shown explicitly.

The Eq.\mref{11} describes the time evolution of $\chi^\sigma(t,s_1,...,s_{n-1},\lambda)$ along trajectories starting
on the hypersurface $\Sigma_{n-1}$ if its "initial" values on this surface, i.e.
$\chi^\sigma(0,s_1,...,s_{n-1},\lambda)\equiv\chi^\sigma(s_1,...,s_{n-1},\lambda)$ are given.

We are going to consider the equation \mref{11} in the semiclassical limit $\lambda\to+\infty$ looking for its solutions in
the form of the following asymptotic series:
\be
E-E_0=\sum_{k\geq 1}E_k\lambda^{-k-1}\nn\\
\chi^\sigma(t,s_1,...,s_{n-1},\lambda)=\sum_{k\geq 0}\chi_k^\sigma(t,s_1,...,s_{n-1})\lambda^{-k}\nn\\
\chi^\sigma(s_1,...,s_{n-1},\lambda)=\sum_{k\geq 0}\chi_k^\sigma(s_1,...,s_{n-1})\lambda^{-k}
\label{3a}
\ee

Putting $\lambda=1$ in \mref{1}, \mref{2} and \mref{3a} we get approximate semiclassical solutions to the energy
eigenvalue problem of the Schr\"odinger equation.

It is to be noticed that for the selfconsistency reasons the semiclassical series for the energy parameter in
\mref{3a} starts from
the second power of $\lambda^{-1}$, i.e. this ensures the proper hierarchy of steps in the algorithm of semiclassical
calculations by which the higher order terms of the series in \mref{3a} are determined by the lower order ones.

It should be noticed also that despite the fact that $E_0$ enters the classical equation of motion \mref{3} it is still
quantum, i.e. its value depends on $\hbar$ which is considered to have the definite numerical value, i.e. $\hbar$ is not
a parameter. In particular the series \mref{3a} represent the inverse power hierarchy in the formal parameter $\lambda$,
i.e. not in powers of $\hbar$, between subsequent terms.

Moreover $E_0$ if quantized can depend on $\lambda$. However, whatever is this dependence the semiclassical series of
the difference $E-E_0$ must be given by \mref{3a}.

Needless to say the introducing $\lambda$ makes a treatment of the Schr\"odinger equation equivalent of course
to considering it in the limit $\hbar\to 0$, i.e.
semiclassically, clearly however separating the role of $\hbar$ as a parameter from its role defining the microscale
of quantum phenomena.

Substituting \mref{3a} into \mref{11} we get:
\be
\frac{\p\chi_0^\sigma(t,s_1,...,s_{n-1})}{\p t}=0\nn\\
\frac{\p\chi_{k+1}^\sigma(t,s_1,...,s_{n-1})}{\p t}=\nn\\
\frac{\sigma i}{2}\ll(J^{\fr}\triangle\ll(J^{-\fr}\chi_{k}^\sigma(t,s_1,...,s_{n-1})\r)+
2\sum_{l=0}^kE_{k-l+1}\chi_l^\sigma(t,s_1,...,s_{n-1})\r)\nn\\
k=0,1,2,...,
\label{12}
\ee
with the obvious solutions:
\be
\chi_0^\sigma(t,s_1,...,s_{n-1})\equiv\chi_0^\sigma(s_1,...,s_{n-1})\nn\\
\chi_{k+1}^\sigma(t,s_1,...,s_{n-1})=\chi_{k+1}^\sigma(s_1,...,s_{n-1})+\nn\\
\frac{\sigma i}{2}\int_0^t\ll(J^{\fr}\triangle\ll(J^{-\fr}\chi_{k}^\sigma(t',s_1,...,s_{n-1})\r)+
2\sum_{l=0}^kE_{k-l+1}\chi_l^\sigma(t',s_1,...,s_{n-1})\r)dt'\nn\\
k=0,1,2,...,
\label{12z}
\ee

\section{Semiclassical wave function expansion elements as classical objects}

\hskip+2em The recurrent system of equations \mref{12} can be considered also from the classical point of view as defining the time evolutions
of $\chi_k^\sigma(t,s_1,...,s_{n-1}),\;k=0,1,2,...,$ along the classical trajectories ${\bf r}(t)=
{\bar{\bf f}}(t,s_1,...,s_{n-1})$. Namely, for each given initial point ${\bf r}_0(s_1,...,s_{n-1})$ and a given trajectory
emerging from it let $d$ denotes a distance measured along the trajectory from the point ${\bf r}_0(s_1,...,s_{n-1})$ to
the point $(x_1,...,x_n)$ lying on this
trajectory. The set $(d,s_1,...,s_{n-1})$ can be used as the new coordinates instead of $(x_1,...,x_n)$. Their mutual
relations are given by:
\be
d={\bar d}(t,s_1,...,s_{n-1})=\int_0^t\ll|\frac{\p{\bar{\bf f}}(t',s_1,...,s_{n-1})}{\p t'}\r|dt'
\label{12a}
\ee
and
\be
{\bf r}=(x_1,...,x_n)={\bar{\bf f}}({\bar t}(d,s_1,...,s_{n-1}),s_1,...,s_{n-1})\equiv {\bf g}(d,s_1,...,s_{n-1})
\label{12b}
\ee
where $t={\bar t}(d,s_1,...,s_{n-1})$ is the solution of $d={\bar d}(t,s_1,...,s_{n-1})$ with respect to $t$.

Considered on the Lagrange manifold $\Lambda_n$ the system is then governed by the Hamiltonian:
\be
{\bar H}_{\Lambda_n}=\fr\frac{p_d^2}{\ll|\frac{\p{\bf g}(d,s_1,...,s_{n-1})}{\p d}\r|^2}+{\bar V}(d,s_1,...,s_{n-1})
\label{12d}
\ee
where ${\bar V}(d,s_1,...,s_{n-1})\equiv V({\bar{\bf f}}({\bar t}(d,s_1,...,s_{n-1}),s_1,...,s_{n-1}))$ and $p_d$ is
the momentum conjugated with $d$.

The corresponding classical equation of motion for ${\chi}^\sigma(t,s_1,...,s_{n-1},\lambda)$ on the manifold
$\Lambda_n$ is therefore:
\be
\frac{d\chi^\sigma(t,s_1,...,s_{n-1},\lambda)}{d t}=\{\chi^\sigma,{\bar H}_{\Lambda_n}\}+
\frac{\p\chi^\sigma(t,s_1,...,s_{n-1},\lambda)}{\p t}=\frac{\p\chi^\sigma(t,s_1,...,s_{n-1},\lambda)}{\p t}
\label{12e}
\ee
since $\chi^\sigma(t,s_1,...,s_{n-1},\lambda)$ being independent of $d$ commutes with the hamiltonian
${\bar H}_{\Lambda_n}$.

Therefore, from \mref{3a} and \mref{12} we get on the Lagrange manifold $\Lambda_n$:
\be
\frac{d{\chi}_0^\sigma(t,s_1,...,s_{n-1})}{d t}=0\nn\\
\frac{d{\chi}_{k+1}^\sigma(t,s_1,...,s_{n-1})}{dt}=
\frac{\sigma i}{2}J^{\fr}\triangle\ll(J^{-\fr}{\chi}_{k}^\sigma(t,s_1,...,s_{n-1})\r)+
\sigma i\sum_{l=0}^kE_{k-l+1}{\chi}_l^\sigma(t,s_1,...,s_{n-1})\nn\\
k=0,1,2,...,\;\;\;\;\;\;\;\;\;
\label{12f}
\ee

Of course \mref{12z} provides the solution to the equations \mref{12f}.

To conclude $\chi_0^\sigma(t,s_1,...,s_{n-1})$ is constant on the Lagrange manifold $\Lambda_n$ according to \mref{12z} being obviously
independent also of $d$.

It is to be noticed that defining on $\Sigma_{n-1}$ another Lagrange manifold $\Lambda_n'$ by choosing a different set
of trajectories we
obtain also another SWF defined on this new Lagrange manifold, i.e. different from the one defined on $\Lambda_n$. This
is why
the corresponding $\chi_0^\sigma(t,s_1,...,s_{n-1})$ cannot be considered as the global integral of motion, i.e. in the
whole phace space $R_{2n}$ of the Hamiltonian $H$ - each particular SWF is defined only on a particular Lagrange manifold
$\Lambda_n$ corresponding to it.

For future applications it is worth to note that if \mref{11} is obviously invariant on a reparametri\-zation of the hypersurface $\Sigma_{n-1}$
it is also invariant on the following change of variables:
\be
t\to \tau(s_1,...,s_{n-1})-t\nn\\
s_k\to h_k(s_1,...,s_{n-1})\nn\\
k=1,...,n-1
\label{13a}
\ee
if it is accompanied simultaneously by the transformations:
\be
\chi^\sigma(t,s_1,...,s_{n-1},\lambda)\to\nn\\{\chi}^{-\sigma}(t,s_1,...,s_{n-1},\lambda)\equiv
J_1^{-\fr}(h_1^{-1}(s_1,...,s_{n-1}),...,h_{n-1}^{-1}(s_1,...,s_{n-1}))\times\nn\\
\chi^\sigma(\tau(s_1,...,s_{n-1})-t,h_1^{-1}(s_1,...,s_{n-1}),...,h_{n-1}^{-1}(s_1,...,s_{n-1}),\lambda)
\label{13b}
\ee
where $J_1(s_1,...,s_{n-1})$ is the Jacobean of the transformation \mref{13a}.

\section{Semiclassical wave functions in the billiards}

\hspace{15pt}Let us apply the above formalism to construct continuous semiclassical wave functions inside billiards $B$
(Fig.1) vanishing on its boundary. Such a construction will be done in several steps the first one consisting of
some geometrical preliminaries describing a notion of a skeleton, i.e. the closed set  of families of trajectories which
forms a base on which SWF's are constructed.

\subsection{Ray bundles and bundle skeletons}

\hspace{15pt}We shall assume that the billiards is classical, i.e. its boundary $\p B$ is a
\begin{figure}
\begin{center}
\psfig{figure=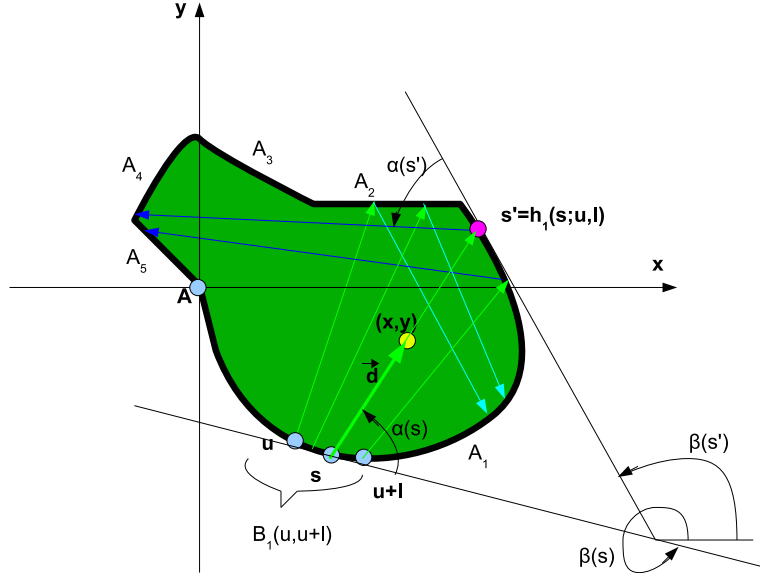,width=12cm}
\caption{An arbitrary billiards}
\end{center}
\end{figure}
closed curve independent of $\lambda$ and given by
${\bf r}={\bf r}_0(s)=[x_0(s),y_0(s)]$ where $s$
is a distance of a boundary point ${\bf r}_0(s)$ measured clockwise along $\p B$ from some other point $A$ of $\p B$ chosen arbitrary, i.e.
$s(A)=0$. Both $x_0(s)$ and $y_0(s)$ are continues. The curve however consists
of a finite number $q,\;q\geq 1$ of smooth arcs $A_1,...,A_q$ with respective length $L_1,...,L_q$, so that the
derivatives $x_0'(s)$ and $y_0'(s)$ are discontinuous
in a finite number of points on the segment $0\leq s\leq L$ where $L=L_1+\cdots+L_q$ is the total length of $\p B$. Both
$x_0(s)$ and $y_0(s)$ are of course periodic with the period equal to $L$. We shall identify the point $A$ with the
point beginning the arc $A_1$.

Next we define a bundle of rays as a family of trajectories in the following way.

Let $A_k(u,l),\;L_1+\cdots+L_{k-1}\leq u\leq L_1+\cdots+L_k,\;0<l\leq L_k,$ be an open connected piece of the
arc $A_k$ beginning at $s=u$ and having a length $l$.

Let further ${\bf r}_k(s,t;u,l),\;u<s<u+l,\;0\leq t,$ be a family of trajectories given by
angles $\gamma_k(s;u,l),\;0\leq\gamma_k(s;u,l)\leq 2\pi$, at which the trajectories escape from  $A_k(u,l)$. The angles
$\gamma(s;u,l)$ are smooth functions of $s$ and are measured with respect to the $x$-axis while the tangential vectors
${\bf t}(s)=[\frac{dx_0(s)}{ds},\frac{dy_0(s)}{ds}]=[\cos\beta(s),\sin\beta(s)]$ are inclined to the $x$-axis at
angles $\beta(s)$ (Fig.1). The latter angle can be discontinuous at the points where $x_0'(s)$ and $y_0'(s)$ are
discontinuous. Then the angle $\alpha_k(s;u,l)=\gamma_k(s;u,l)-\beta_k(s)$ is made by
the classical ball momentum ${\bf p}(s;u,l)$ on the trajectory with the tangent vector ${\bf t}(s)$, i.e.
${\bf p}(s;u,l)\cdot{\bf t}(s)=p\cos\alpha_k(s;u,l)$. It is assumed that $0<\alpha_k(s;u,l)<\pi$.

The classical time evolution of the family ${\bf r}_k(s,t;u,l),\;u<s<u+l,$ is therefore the following
\be
{\bf r}_k(s,t;u,l)={\bf r}_0(s)+{\bf p}(s;u,l)t,\;\;\;\;\;\;{\bf r}_0(s)\in A_k
\label{15}
\ee
where ${\bf p}(s;u,l)=[p\cos\gamma_k(s;u,l),p\sin\gamma_k(s;u,l)]$ satisfies the classical equations of motions
\mref{3}, i.e. ${\bf p}^2(s;u,l)=2E_0$ (again we put $m=1$ for the billiard ball mass).

The trajectories \mref{15} define of course the change of variables $(x,y)\to(t,s)$, in vicinity of $A_k(u,l)$, i.e.
$x=f_k(t,s;u,l),\;y=g_k(t,s;u,l)$ with the Jacobean:
\be
{\tilde J}_k(t,s;u,l)=p^2\gamma_k'(s;u,l)t-p\ll|{\bf t}(s)\r|\sin\alpha_k(s;u,l)=
p^2\gamma_k'(s;u,l)t-p\sin\alpha_k(s;u,l)\nn\\
\label{16}
\ee
since $\ll|{\bf t}(s)\r|=1$.

The family of trajectories defined in the above way will be called a {\bf bundle of rays} emerging from the segment
$A_k(u,l)$ of $A_k$ and will be denoted by $B_k(u,l)$ while the trajectory itself will be called {\bf rays}.

Since each ray of the bundle $B_k(u,l)$ after some time $\tau_k(s;u,l),\;{\bf r}_0(s)\in A_k(u,l)$, (different for
different rays)
achieves another point of the boundary $\p B$ it means that the bundle $B_k(u,l)$ maps the segment $A_k(u,l)$ into
another piece $BA_k(u,l)$ of the boundary $\p B$. In general it is assumed that this map of $A_k(u,l)$ into
$BA_k(u,l)$ provided
by the transformation \mref{15} is one-to-one except the caustic points of $BA_k(u,l)$ in which
${\tilde J}_K(\tau_k(s;u,l),h_k(s;u,l);u,l)=0,\;{\bf r}_0(h_k(s;u,l))\in BA_k(u,l)$. Here $h_k(s;u,l),\;
{\bf r}_0(s)\in
A_k(u,l)$, realizes explicitly this map. The caustic points will be assumed to be isolated singular points of the map
$A_k(u,l)\to BA_k(u,l)$. Typically they need a special care if they are.

By $DB_k(u,l)$ will be denoted a domain of the billiards $B$ covered by rays of the bundle $B_k(u,l)$ emerging from
$A_k(u,l)$ and ending at $BA_k(u,l)$. The domain $DB_k(u,l)$ is locally a Lagrangian manifold on which each loop
integral $\oint{\bf p}\cdot d{\bf r}$ vanishes.

Assume further the boundary $\p B$ to be a mirror-like, i.e. reflecting the incoming rays according to the reflection
principle of the geometrical optics and let $A_{k'}(u',l')$ be a piece of another arc $A_{k'}$ of $\p B$ such that
$A_{k'}(u',l')\cap BA_k(u,l)\neq\oslash$ on which another bundle of rays
$B_{k'}(u',l')=\{{\bf r}_{k'}(s,t;u',l'):u'<s<u'+l',\;0\leq t\}$ is defined.

If on the segment $A_{k'}(u',l')\cap BA_{k}(u,l)$ the ray bundle $B_{k'}(u',l')$ coincides with the reflected one we
call the ray bundle $B_{k'}(u',l')$ a reflection of the bundle $B_k(u,l)$ on the segment mentioned.

The reflection operation over the bundle $B_k(u,l)$ will be denoted by $\Pi$ so that $\Pi B_k(u,l)$ denotes the set of
all rays arising by the reflection of the bundle $B_k(u,l)$ of $\p B$.

Consider now a family of the disjoint ray bundles ${\bf B}=\bigcup B_k(u,l),\;B_k(u,l)\cap B_{k'}(u',l')=
\oslash$ if $B_k(u,l)\neq B_{k'}(u',l')$.

The family {\bf B} will be called closed under reflection $\Pi$ on the boundary $\p B$ if the following two conditions
are satisfied for each bundle $B_k(u,l)\in{\bf B}$:
\be
\Pi B_k(u,l)=\bigcup_{j=1}^n B_j(u_j,l_j)\cap \Pi B_k(u,l),\;\;\;\;\;\;\;B_j(u_j,l_j)\in A_{i_j},\;j=1,...,n\nn\\
B_k(u,l)=\bigcup_{i=1}^m \Pi B_i(u_i,l_i)\cap B_k(u,l),\;\;\;\;\;\;\;B_i(u_i,l_i)\in A_{j_i},\;i=1,...,m\nn\\
\Pi{\bf B}={\bf B}
\label{16a}
\ee

A closed family ${\bf B}'=\bigcup B_{k'}(u',l')$ is embedded into a closed family ${\bf B}=\bigcup B_k(u,l)$ if each ray
bundle of ${\bf B}'$ is a subset of some ray bundle of ${\bf B}$ and each bundle of ${\bf B}$ contains some bundle of
${\bf B}'$.

A closed bundle family is called connected if it cannot be decomposed into a sum of another two disjoint closed
bundle families.

A closed connected bundle family will be called a {\bf Lagrange bundle skeleton} or simply a {\bf skeleton} if it cannot
be embedded into another closed connected bundle family.

From now on all considered bundle families will be assumed to be skeletons.

A basic property of each ray belonging to a skeleton ${\bf B}$ is that by its time evolution and bounces on
the billiards boundary it will never leave ${\bf B}$.

Another basic property of a skeleton ${\bf B}$ is its completeness, i.e. none bundle can be added to or removed from
the skeleton not destroying it.

Let us now reverse in time all trajectories belonging to ${\bf B}$. This operation leads us again to some skeleton
${\bf B}^A$ which will be called associated with ${\bf B}$.

Bundles of ${\bf B}^A$ are obtained simply from the corresponding bundles
of ${\bf B}$. Namely with each bundle $B_k(u,l)$ of ${\bf B}$ let us associate a bundle $B_k^A(u,l)$ which
trajectories satisfy the following condition:
\be
\gamma_k^A(s;u,l)=\pi+2\beta(s)-\gamma_k(s;u,l),\;\;\;\;u<s<u+l
\label{16b}
\ee
i.e. these trajectories are just the reflections on $\p B$ of the time reversed trajectories defined by
$\gamma_k(s;u,l)$ and belonging to $B_k(u,l)$.

The skeleton ${\bf B}^A$ is organized by all bundles $B_k^A(u,l)$ obtained in the above way.

\subsection{SWF's defined on ray bundles of a bundle skeleton}

\hspace{15pt}Consider a bundle skeleton ${\bf B}$. On each of its ray bundle $B_k(u,l)$ we can now define the following
pair of SWF's $\Psi_k^\sigma(t,s;u,l;\lambda),\;\sigma=\pm$:
\be
\Psi_k^\sigma(t,s;u,l;\lambda)={\tilde J}_k^{-\fr}(t,s;u,l)
e^{\sigma i\lambda\ll(p^2t+p\int_0^s\cos\alpha_k(s;u,l)ds'\r)}\chi_k^\sigma(t,s;u,l;\lambda)
\label{17}
\ee
where $p^2=2E_0$ and $\chi_k^\sigma(t,s;u,l;\lambda),\;\sigma=\pm$, are given by \mref{3a}
and \mref{12z}.

Exactly in the same way we can define a pair $\Psi_{A;k}^\sigma(t,s;u,l;\lambda),\;\sigma=\pm$, of SWF's on the
corresponding associated bundle $B_k^A(u,l)$:
\be
\Psi_{A;k}^\sigma(t,s;u,l;\lambda)={\tilde J}_{A;k}^{-\fr}(t,s;u,l)
e^{\sigma i\lambda\ll(p^2t-p\int_0^s\cos\alpha_k(s;u,l)ds'\r)}\chi_{A;k}^\sigma(t,s;u,l;\lambda)
\label{17a}
\ee

It will be also convenient for further considerations to substitute the time variable $t$ by the distance variable
$d=pt$ and consequently to give the trajectories \mref{15}, the Jacobean \mref{16}
and the wave function \mref{17} the following forms:
\be
{\bf r}_k(d,s;u,l)={\bf r}_0(s)+{\bf d}(s;u,l)\nn\\
{\bf d}(s;u,l)={\bf p}t=[d\cos\gamma_k(s;u,l),d\sin\gamma_k(s;u,l)]\nn\\
{\bf r}_0(s)\in A_k(u,l)
\label{15a}
\ee
and
\be
J_k(d,s;u,l)=\frac{1}{p}{\tilde J}_k(t,s;u,l)=\frac{\p\gamma_k(s;u,l)}{\p s}d-\sin\alpha_k(s;u,l)\nn\\
{\bf r}_0(s)\in A_k(u,l)
\label{16c}
\ee
and
\be
\Psi_k^\sigma(d,s;u,l;\lambda)=J_k^{-\fr}(d,s;u,l)
e^{\sigma ik\ll(d+\int_0^s\cos\alpha_k(s';u,l)ds'\r)}{\bar\chi}_k^\sigma(d,s;u,l;\lambda)\nn\\
{\bf r}_0(s)\in A_k(u,l)
\label{17A}
\ee
where ${\bar\chi}_k^\sigma(d,s;u,l;\lambda)\equiv p\chi_k^\sigma(\frac{d}{p},s;u,l;\lambda),\;\sigma=\pm$, and
$k=p\lambda$ is the wave number of the billiards ball.

Nevertheless, for simplicity of notations, the bar over ${\bar\chi}_k^\sigma(d,s;u,l;\lambda)$ will be dropped in our
further considerations.

By the variable $d$ the solutions \mref{12z} can be rewritten in the form:
\be
{\chi}_{k,0}^\sigma(d,s;u,l)\equiv{\chi}_{k,0}^\sigma(s;u,l)\nn\\
{\chi}_{k,j+1}^\sigma(d,s;u,l)={\chi}_{k,j+1}^\sigma(s;u,l)+\nn\\
\frac{\sigma i}{2p}\int_0^d\ll(D_k(a,s;u,l){\chi}_{k,j}^\sigma(a,s;u,l)+
2\sum_{m=0}^jE_{j-m+1}{\chi}_{k,m}^\sigma(a,s;u,l)\r)da\nn\\
j=0,1,2,...,
\label{13A}
\ee
where $D_k(d,s;u,l)=J_k^\fr(d,s;u,l)\cdot\triangle_k(d,s;u,l)\cdot J_k^{-\fr}(d,s;u,l)$ and $\triangle_k(d,s;u,l)$ is
the Laplacean expressed by the variables $d$ and $s$ corresponding to the $B_k(u,l)$-bundle.

$\Psi_k^\sigma(d,s;u,l;\lambda),\;\sigma=\pm,$ are defined initially in the domain $D_k(u,l)$ of the billiards
which boundary $\p D_k(u,l)$ contains of course $A_k(u,l)$. A pattern of the remaining part of $\p D_k(u,l)$ depends on the corresponding
caustic $K_k(u,l)=\{(f_k(K_k(s;u,l),s;u,l),g_k(K_k(s;u,l),s;u,l)):$\linebreak$J_k(K_k(s;u,l),s;u,l)=0,\; {\bf r}_0(s)
\in A_k(u,l)\}$.
It follows from \mref{16} that
\be
K_k(s;u,l)=\frac{\sin\alpha_k(s;u,l)}{\frac{\p\gamma_k(s;u,l)}{\p s}},\;\;\;\;\;\;\; {\bf r}_0(s)\in A_k(u,l)
\label{18}
\ee
and $K_k(s;u,l)$ can escape to infinity if $\frac{\p\gamma_k(s;u,l)}{\p s}\to 0$.

If the set $K_k(u,l)\cap B$ is not empty then it belongs to $\p D_k(u,l)$. The remaining rays of $B_k(u,l)$ which are not
tangent to $K_k(u,l)\cap B$ end at $BA_k(u,l)$ and complete in this way the domain $D_k(u,l)$.

An important property of the representation \mref{17} is its uniqueness, i.e. for two
different bundles defined on the segment $A_k(u,l)$
this representation provides us with two different pairs of $\Psi_k^\sigma(d,s;u,l;\lambda),\;\sigma=\pm$. This conclusion follows
from the fact that for $\lambda$ sufficiently large the SWF's are determined only by exponentials and the latter are
different at the same points $(x,y)$ for different bundles.

\subsection{Propagation of $\Psi_k^\sigma(d,s;u,l;\lambda)$ through a skeleton}

\hspace{15pt}$\Psi_k^\sigma(d,s;u,l;\lambda),\;\sigma=\pm,$ being defined in $D_k(u,l)\in B_k(u,l)$ can be continued on
the whole skeleton by accepting two rules the solutions have to satisfy.

A need for the first rule to be formulated arises when the solutions propagate by the caustic where they are singular. These are
branch points singularities on the $d$-plane of the type $(d-d_c(s))^{-\fr}$ where $J_k(d_c(s),s;u,l)=0$. These points have to
be avoided somehow when continuing the solutions along the rays of $B_k(u,l)$. Relying on the results of App.B we accept the rule
of avoiding these points clockwise by $\Psi_k^+(d,s;u,l;\lambda)$ and
above them and anticlockwise by $\Psi_k^-(d,s;u,l;\lambda)$ and below them.

The second rule has to be given to describe bouncings of $\Psi_k^\sigma(d,s;u,l;\lambda),\;\sigma=\pm,$ by the billiards
boundary.

Let $B_{k'}(u',l')$ be a reflection of the bundle $B_k(u,l)$ i.e. $B_{k'}(u',l')\subset\Pi B_k(u,l)$. Continuations of
$\Psi_k^\sigma(d,s;u,l;\lambda),\;\sigma=\pm,$ onto $B_{k'}(u',l')$ is given by the following formula:
\be
\Psi_k^{\sigma,cont}(d,s;u',l';\lambda)=-\eta_\sigma J_{k'}^{-\fr}(d,s;u',l')\times\nn\\
e^{\sigma i\lambda p\ll(d-\int_{u'}^s\cos\alpha_{k'}(s';u',l')ds'+\delta_k(u,l)\r)}
\chi_k^{\sigma,cont}(d,s;u',l';\lambda)\nn\\
\delta_k(u,l)=D(h_k^{-1}(s;u,l);u,l)+\int_{u'}^s\cos\alpha_{k'}(s';u',l')ds'-\int_u^{h_k^{-1}(s;u,l)}\cos\alpha_k(s';u,l)ds'\nn\\
\chi_k^{\sigma,cont}(0,s;u',l';\lambda)=\ll|\frac{\p h_k(s';u,l)}{\p s'}\r|_{s'=h_k^{-1}(s;u,l)}^{-\fr}
\chi_{k}^{\sigma}(D(h_k^{-1}(s;u,l);u,l),h_k^{-1}(s;u,l);u,l;\lambda)\nn\\
{\bf r}_0(s)={\bf r}_0(h_k^{-1}(s;u,l))+{\bf D}(h_k^{-1}(s;u,l);u,l)
\label{18a}
\ee

The factor $\eta_\sigma$ in \mref{18a} depends on whether there was a caustic point on the way of continuations of
$\Psi_k^{\sigma}(d,s;u,l;\lambda)$ in which case $\eta_\sigma=\sigma i$ but $\eta_\sigma\equiv 1$ in the
opposite case.

$D(s;u,l)=|{\bf D}(s;u,l)|$ denotes the distance between the points ${\bf r}_0(h_k(s;u,l))$ and
${\bf r}_0(s)$ of the boundary $\p B$.

It is to be noticed that $\delta_k(u,l)$ is $s$-independent (see App.C).

$\chi_k^{\sigma,cont}(d,s;u',l';\lambda)$ is defined by \mref{13A} with the initial values given by the corresponding
semiclassical series of $\chi_k^{\sigma,cont}(0,s;u',l';\lambda)$ as given by \mref{18a}. Due to the chain property of
the formulae \mref{13A} such a definition of $\chi_k^{\sigma,cont}(d,s;u',l';\lambda)$ ensures that
$\chi_k^\sigma(d,s;u,l;\lambda)$ can be considered as a continuous function of $d$ on the skeleton as $d$ changes along
rays arising by subsequent reflections of the one starting from the bundle $B_k(u,l)$.

Another property of the definition \mref{18a} is that close to the arc $A_{k'}$ there is a domain in which both the SWF's
$\Psi_k^{\sigma,cont}(d',s';u',l';\lambda)$ and $\Psi_k^\sigma(d,s;u,l;\lambda)$ are defined so that
the superposition:
\be
\Psi_k^{\sigma,sup}(x,y,\lambda)\equiv\Psi_k^{\sigma,cont}(d',s';u',l';\lambda)+\Psi_k^\sigma(d,s;u,l;\lambda)\nn\\
x=x_0(s)+d\cos\alpha_k(s;u,l)=x_0(s')+d'\cos\alpha_{k'}(s';u',l')\nn\\
y=y_0(s)+d\sin\alpha_k(s;u,l)=y_0(s')+d'\sin\alpha_{k'}(s';u',l')
\label{18b}
\ee
can be done.

It follows easily from \mref{18a} that on the arc $A_{k'}$, i.e. for $d'=0$ this superposition vanishes.

\subsection{SWF's vanishing on the billiards boundary}

\hspace{15pt}Another obvious property of $\Psi_k^\sigma(d,s;u,l;\lambda),\;\sigma=\pm$, is that they cannot vanish on $A_k(u,l)$
unless $\chi_k^\sigma(d,s;u,l;\lambda)$ vanish there
identically. Therefore a wave function $\Psi_k^{as;\sigma}(x,y;u,l;\lambda)$ vanishing on $A_k(u,l)$ should be represented in the
semiclassical limit by a linear combination of at least two SWF's of the form \mref{17}. The superposition \mref{18b} is an important
example of such a combination. It is shown however in App.A on a more general level of considerations that the
proper linear combinations have to be the following:
\be
\Psi_k^{as;\sigma}(x,y;u,l;\lambda)=\Psi_{k}^{\sigma}(d_1,s_1;u,l;\lambda)+\Psi_{A;k}^{-\sigma}(d_2,s_2;u,l;\lambda)=\nn\\
J_{k}^{-\fr}(d_1,s_1;u,l)e^{i\sigma k\ll(d_1+\int_0^{s_1}\cos\alpha_{k}(s';u,l)ds'\r)}
\chi_{k}^{\sigma}(d_1,s_1;u,l;\lambda)+\nn\\
J_{A;k}^{-\fr}(d_2,s_2;u,l)e^{-i\sigma k\ll(d_2-\int_0^{s_2}\cos\alpha_{k}(s';u,l)ds'\r)}
\chi_{A;k}^{-\sigma}(d_2,s_2;u,l;\lambda)
\label{19}
\ee
with the following boundary conditions:
\be
\chi_{k}^{\sigma}(0,s;u,l;\lambda)+\chi_{A;k}^{-\sigma}(0,s;u,l;\lambda)=0
\;\;\;\;\;\;\; {\bf r}_0(s)\in A_k(u,l)
\label{19a}
\ee
while the point $(x,y)$ is the cross point of the respective trajectories belonging to different bundles, i.e.
\be
{\bf r}\equiv[x,y]={\bf r}_{k}(d_1,s_1;u,l)={\bf r}_0(s_1)+{\bf d}_{1}(s_1;u,l)=\nn\\
{\bf r}_{A;k}(d_2,s_2;u,l)={\bf r}_0(s_2)+{\bf d}_{2}(s_2;u,l)\nn\\
{\bf r}_{k}(d,s;u,l)\in B_k(u,l),\;\;\;\;\;{\bf r}_{A;k}(d,s;u,l)\in B_k^A(u,l)
\label{20}
\ee

The vanishing superposition \mref{19} if defined on the bundle $B_{k'}(u',l')$ suggests that
\linebreak $\Psi_{A;k'}^{-\sigma}(d,s;u',l';\lambda)$ should be related somehow to the SWF
$\Psi_k^{\sigma}(d,h_k^{-1}(s;u,l);u,l);u,l;\lambda)$ defined on the bundle $B_{k}(u,l)$ which the previous one is
a reflection. In the next section this relation is established as a condition matching both the solutions.

\subsection{Matching solutions defined on two different bundles of a skeleton}

\hspace{15pt}Let now $\Psi_k^{as;\sigma}(x,y;u,l;\lambda)$'s be defined in the above way on the corresponding bundles of {\bf B} and
${\bf B}^A$. Suppose that the corresponding
SWF's $\Psi_k^{\sigma}(d,s;u,l;\lambda),\;\sigma=\pm,$ and $\Psi_k^{A;\sigma}(d,s;u,l;\lambda),\;\sigma=\pm,$ which
construct them can be continued along the respective bundles on
which they are defined. Note that the domains $D_k(u,l)$ and $D_k^A(u,l)$ where these SWF's are initially defined are then
extended to $DB_k(u,l)$ and $DB_k^A(u,l)$ respectively. We are then faced with the problem of matching the continued
solutions with the others defined on the same skeletons {\bf B} and ${\bf B}^A$.

Anticipating the results of App.B such a matching should be done as follows.

Let $B_k(u,l)$ be a reflection of the bundles $B_{k_1}(u_1,l_1),B_{k_2}(u_2,l_2),\ldots,B_{k_n}(u_n,l_n)$ satisfying \mref{16a}.
Let $\Psi_{k_1}^{\sigma}(d,s;u_1,l_1;\lambda),\Psi_{k_2}^{\sigma}(d,s;u_2,l_2;\lambda),\ldots,\Psi_{k_n}^{\sigma}(d,s;u_n,l_n;
\lambda)$ be defined and continued on the respective bundles $B_{k_1}(u_1,l_1),\;B_{k_2}(u_2,l_2),\;\ldots,B_{k_n}(u_n,l_n)$. Let further
\linebreak $\Psi_{A;k}^{-\sigma}(d,s;u,l;\lambda)$ be defined on the bundle $B_k^A(u,l)$ while
$\Psi_k^{\sigma}(d,s;u,l;\lambda)$ on the bundle $B_k(u,l)$ being both related by the boundary condition \mref{19a}.

Then we make the following identification:
\be
\Psi_{A;k}^{-\sigma}(d,h(s;u_j,l_j);u,l;\lambda)=\Psi_{k_j}^{\sigma}(D(s;u_j,l_j)-d,s;u_j,l_j;\lambda)\nn\\
{\bf r}_0(h(s;u_j,l_j))\in A_k(u,l)\cap BL(u_j,l_j)\nn\\
{\bf r}_0(s)\in L(u_j,l_j)\nn\\
{\bf r}_0(h(s;u_j,l_j))={\bf r}_0(s)+{\bf D}(s;u_j,l_j),\;\;\;\;\;j=1,...n
\label{22}
\ee

Rewritten in terms of the $\chi$-coefficients eq.\mref{22} gives:
\be
\chi_{A;k}^{-\sigma}(d,h(s;u_j,l_j);u,l;\lambda)=\nn\\
\eta_\sigma e^{\sigma i\lambda p\delta(u_j,l_j)}
\ll|\frac{\p h(s;u_j,l_j)}{\p s}\r|^{-\fr}\chi_{k_j}^{\sigma}(D(s;u_j,l_j)-d,s;u_j,l_j;\lambda)\nn\\
\delta(u_j,l_j)=D(s;u_j,l_j)+\int_{u_j}^s\cos\alpha_{k_j}(s';u_j,l_j)ds'-\int_u^{h(s;u_j,l_j)}\cos\alpha_k(s';u,l)ds'\nn\\
{\bf r}_0(h(s;u_j,l_j))\in A_k(u,l)\cap BL(u_j,l_j)\nn\\
{\bf r}_0(s)\in L(u_j,l_j)\nn\\
{\bf r}_0(h(s;u_j,l_j))={\bf r}_0(s)+{\bf D}(s;u_j,l_j),\;\;\;\;\;j=1,...n
\label{22a}
\ee

As previously $\delta(u_j,l_j)$ in the above formula is $s$-independent (see App.C). Due to that and due to the
properties \mref{13a} and \mref{13b} the rhs of \mref{22a} satisfies \mref{11} as it should.

Putting $d=0$ in \mref{22a} and taking into account \mref{19a} we get:
\be
\chi_k^{\sigma}(0,h(s;u_j,l_j);u,l;\lambda)=-\chi_{A;k}^{-\sigma}(0,h(s;u_j,l_j);u,l;\lambda)=\nn\\
-\eta_\sigma e^{\sigma i\lambda p\delta(u_j,l_j)}\ll|\frac{\p h(s;u_j,l_j)}{\p s}\r|^{-\fr}
\chi_{k_j}^{\sigma}(D(s;u_j,l_j),s;u_j,l_j;\lambda)=\nn\\
-\eta_\sigma e^{\sigma i\lambda p\delta(u_j,l_j)}\chi_{k_j}^{\sigma,cont}(0,h(s;u_j,l_j);u,l;\lambda)\nn\\
{\bf r}_0(h(s;u_j,l_j))\in A_k(u,l)\cap BL(u_j,l_j)\nn\\
{\bf r}_0(s)\in L(u_j,l_j)\nn\\
{\bf r}_0(h(s;u_j,l_j))={\bf r}_0(s)+{\bf D}(s;u_j,l_j),\;\;\;\;\;j=1,...n
\label{22b}
\ee

Comparing \mref{22b} with \mref{18a} it is easily seen that $\Psi_k^{\sigma}(d,s;u,l;\lambda)$ has to coincide
with $\Psi_{k_j}^{\sigma,cont}(d,s;u,l;\lambda)$.

\subsection{SWF's defined on a bundle skeleton}

\hspace{15pt}Let ${\bf r}=(x,y)$ be a fixed point of the billiards. Let $D(x,y)$ denote a set of all $D(u,l)$,\- $B(u,l)\in{\bf B}$
which contain this point and $D_A(x,y)$ the respective set of $D^A(u,l),\; B^A(u,l)\in{\bf B}^A$. Let further $DB(x,y)$
denote a set of all $DB(u,l)$,\- $B(u,l)\in{\bf B}$
which contain the point $(x,y)$ and $DB^A(x,y)$ the respective set of $DB^A(u,l),\; B^A(u,l)\in{\bf B}^A$.

SWF's $\Psi_{\bf B}^{as;\sigma}(x,y,\lambda)$ vanishing on the billiards boundary can now be defined on ${\bf B}$ and
${\bf B}^A$ as follows:
\be
\Psi_{\bf B}^{as;\sigma}(x,y,\lambda)=\nn\\\sum_{D(u,l)\in D(x,y)}\Psi^\sigma(d(u,l),s(u,l);u,l;\lambda)+
\sum_{D^A(u,l)\in D_A(x,y)}\Psi_A^{-\sigma}(d(u,l),s(u,l);u,l;\lambda)=\nn\\
\sum_{DB(u,l)\in DB(x,y)}\Psi^\sigma(d(u,l),s(u,l);u,l;\lambda)=
\sum_{DB^A(u,l)\in DB^A(x,y)}\Psi_A^{-\sigma}(d(u,l),s(u,l);u,l;\lambda)\nn\\
{\bf r}={\bf r}_0(s(u,l))+{\bf d}(u,l)
\label{23}
\ee

The SWF's $\Psi^\sigma(d,s;u,l;\lambda)$ defined on bundles of ${\bf B}$ and $\Psi_A^\sigma(d,s;u,l;\lambda)$ defined on
respective bundles of ${\bf B}^A$ are
related with each other by the boundary conditions \mref{19a} and by matching conditions \mref{22}-\mref{22b}.

The solutions $\Psi_{\bf B}^{as;+}(x,y,\lambda)$ and $\Psi_{\bf B}^{as;-}(x,y,\lambda)$ coincide if and only if
${\bf B}={\bf B}^A$.

It is clear that the conditions \mref{22b} have to determine also the $\chi$-factors $\chi_k^\sigma(s;u,l;\lambda)$
for all the bundles $B_k(u,l)$ which are the "initial" conditions for both $\chi_k^\sigma(d,s;u,l;\lambda)$ and
$\chi_{A;k}^\sigma(d,s;u,l;\lambda)$, i.e.
$\chi_k^\sigma(s;u,l;\lambda)\equiv\chi_k^\sigma(0,s;u,l;\lambda)\equiv-\chi_{A;k}^\sigma(0,s;u,l;\lambda)$. Nevertheless these conditions
cannot be given arbitrarily. Just opposite all $\chi_k(s;u,l;\lambda)$ have to satisfy \mref{22b} in selfconsistent
way. It is not easy to predict how this can be realized.

Certainly one has to take into account that
$\chi_k^\sigma(s;u,l;\lambda)$ being given at the point ${\bf r}_0(s(u,l))$ of the bundle $B_k(u,l)$ are distributed along
the boundary $\p B$ by subsequent reflections of the initial ray ${\bf r}={\bf r}_0(s(u,l))+{\bf d}(u,l)$ which carries
on $\chi_k^\sigma(d,s;u,l;\lambda)$ according the formulae \mref{13A}, \mref{18a} and \mref{22b}. However the ray
come back eventually to the segment
$A_k(u,l)$ defining $\chi_k^\sigma(s;u,l;\lambda)$ in some new point ${\bf r}_0(s'(u,l))$ of this segment.
Naturally a value of $\chi_k^{\sigma,cont}(D(u,l),s'(u,l);u',l';\lambda)$ at this point has
to coincide with $\chi_k^\sigma(s'(u,l);u,l;\lambda)$ defined already at this point. This is just one of the selfconsistency conditions
contained in \mref{22b} (see "the last quantization condition" \mref{23c} below).

It seems difficult however to find a general systematic way of exploring the system \mref{22b} to get solutions satisfying it.
Nevertheless one can always think at least minimally trying to guess reasonable solutions to this system. In the next sections
we will try to use both the approaches for some particular cases of billiards.

The formulae \mref{22a} and \mref{22b} defines the conditions which the SWF's
$\chi_k^{\sigma}(d,h(s;u_j,l_j);u,l;\lambda)$ should satisfy when bouncing from the billiards boundary. Nevertheless
this condition should be specified additionally with respect to its factors. Namely, taking their large $\lambda$-limit
we get:
\be
\chi_{k,0}^{\sigma}(h(s;u_j,l_j);u,l)=-\delta\ll|\frac{\p h(s;u_j,l_j)}{\p s}\r|^{-\fr}
e^{\sigma i\lambda p\delta(u_j,l_j)}\chi_{k_j,0}^{\sigma}(s;u_j,l_j)\nn\\
\chi_{k,r+1}^{\sigma}(h(s;u_j,l_j);u,l)=-\delta\ll|\frac{\p h(s;u_j,l_j)}{\p s}\r|^{-\fr}
e^{\sigma i\lambda p\delta(u_j,l_j)}\times\nn\\
\ll(\chi_{k_j,r+1}^{\sigma}(s;u_j,l_j)+\frac{\sigma i}{2p}\int_0^{D(s;u_j,l_j)}\ll(\frac{{}^{}}{{}_{}}J^{\fr}\triangle\ll(J^{-\fr}\chi_{k_j,r}^\sigma(a,s;u_j,l_j)\r)\r.\r.+\nn\\
\ll.\ll.2\sum_{l=0}^rE_{r-l+1}\chi_{k_j,l}^\sigma(a,s;u_j,l_j)\r)da\r)\nn\\
r=0,1,2,...,
\label{22c}
\ee

The above equations should be satisfied on each bundle $B_k(u,l)$ of the skeleton ${\bf B}$.

The first of the equations \mref{22c} should determine the classical quantities, namely the skeleton {\bf B} and the
"classical" energy $E_0=\fr p^2$ and by them define the JWKB approximation of the SWF's. Namely:
\be
\Psi_{\bf B}^{JWKB;\sigma}(x,y,\lambda)=\sum_{DB(u,l)\in D(x,y)}\Psi^{JWKB;\sigma}(d(u,l),s(u,l);u,l)=\nn\\
\sum_{DB(u,l)\in D(x,y)}J^{-\fr}(d(u,l),s(u,l))
e^{i\lambda p\ll(d(u,l)+\int_{u}^{s(u,l)}\cos\alpha(s';u,l)ds'\r)}\chi_0^\sigma(d(u,l),s(u,l);u,l)
\label{22e}
\ee

The remaining equations determine quantum corrections to the "classical" ones involved in \mref{22e}.

However it is easy to note that for the selfconsistency of the equations \mref{22c} it is necessary for the exponent
$e^{\sigma i\lambda p\delta(u_j,l_j)}$ to be independent of $\lambda$, i.e. we have to have on each bundle $B_k(u,l)$
of {\bf B}:
\be
\lambda p\delta_k(u,l)=\phi_k(u,l)\nn\\
B_k(u,l)\subset {\bf B}
\label{22d}
\ee
where $\delta_k(u,l)$ is given by \mref{18a} and $\phi_k(u,l)$ is a $\lambda$-independent constant.

The equations \mref{22d} have to define both the skeleton {\bf B} and the energy $E_0=\fr p^2$.

Taking into account the last conclusions we get the following final set of the recurrent quantization conditions:
\be
\lambda p\delta_k(u,l)=\phi_k(u,l)\nn\\
\chi_{k,0}^{\sigma}(h(s;u_j,l_j);u,l)=-\delta e^{\sigma i\phi_{k_j}(u_j,l_j)}\ll|\frac{\p h(s;u_j,l_j)}{\p s}\r|^{-\fr}\chi_{k_j,0}^{\sigma}(s;u_j,l_j)\nn\\
\chi_{k,r+1}^{\sigma}(h(s;u_j,l_j);u,l)=
-\delta e^{\sigma i\phi_{k_j}(u_j,l_j)}\ll|\frac{\p h(s;u_j,l_j)}{\p s}\r|^{-\fr}
\ll(\frac{{}^{}}{{}_{}}\chi_{k_j,r+1}^{\sigma}(s;u_j,l_j)+\r.\nn\\
\ll.\frac{\sigma i}{2p}\int_0^{D(s;u_j,l_j)}\ll(\frac{{}^{}}{{}_{}}J^{\fr}\triangle\ll(J^{-\fr}\chi_{k_j,r}^\sigma(a,s;u_j,l_j)\r)+
2\sum_{l=0}^rE_{r-l+1}\chi_{k_j,l}^\sigma(a,s;u_j,l_j)\r)da\r)\nn\\
r=0,1,2,...,
\label{22f}
\ee
together with:
\be
\chi_{k,0}^{\sigma}(d,s;u,l)\equiv \chi_{k,0}^{\sigma}(s;u,l)\nn\\
\chi_{k,r+1}^{\sigma}(d,s;u,l)=\chi_{k,r+1}^{\sigma}(s;u,l)+\nn\\
\frac{\sigma i}{2p}\int_0^d
\ll(J^{\fr}\triangle\ll(J^{-\fr}\chi_{k,r}^\sigma(a,s;u,l)\r)+
2\sum_{m=0}^rE_{r-l+1}\chi_{k,m}^\sigma(a,s;u,l)\r)da\nn\\
r=0,1,2,...,
\label{22g}
\ee

\subsection{Continuity of SWF's built on a skeleton}

\hspace{15pt}A skeleton on which SWF $\Psi^{as}(x,y,\lambda)$ is built in the way described above does not ensure
automatically that this SWF will be defined continuously on it. It is certainly continuous inside each bundle of the skeleton but it
can appear to be discontinuous if a point $(x,y)$ crosses bundle boundaries. In general the following circumstances can
accompany such a crossing:
\begin{enumerate}
\item the crossed boundary of $DB_k(u,l)$ coincides with a boundary of $DB_{k'}(u+l,l')$ of a neighboring bundle
$B_{k'}(u+l,l')$;
\item there is no such respective neighboring bundle.
\end{enumerate}

In the first of the above cases the corresponding $\Psi_k^\pm(d,s;u,l;\lambda),\;(x(d,s),y(d,s))\in DB_k(u,l)$, and
$\Psi_{k'}^\pm(d,s;u+l,l';\lambda),\;(x(d,s),y(d,s))\in DB_{k'}(u+l,l')$, defined by \mref{19} have be identified
on the common boundary, i.e.
\be
\Psi_k^\pm(d,s;u,l;\lambda)=\Psi_{k'}^\pm(d,s;u+l,l';\lambda)\nn\\
(x(d,s),y(d,s))\in \p DB_k(u,l)\cap\p DB_{k'}(u+l,l')\neq\oslash
\label{23a}
\ee

In the second case however the corresponding SWF's $\Psi_k^\pm(d,s;u,l;\lambda),\;(x(d,s),y(d,s))\in DB_k(u,l)$
have to vanish on such a boundary of $DB_k(u,l)$, i.e.
\be
\Psi_k^\pm(d,s;u,l;\lambda)=0\nn\\
(x(d,s),y(d,s))\in\p DB_k(u,l)
\label{23b}
\ee

The last condition though necessary seems to be a little bit arbitrary. However we should remember that our
calculations are performed in the semiclassical regime, i.e. in the classically allowed regions (bundles) outside which
the semiclassical wave functions cannot exist. Physically this means obviously that outside each bundle a
corresponding piece of the wave function represented on the bundle by its respective semiclassical approximation has to
vanish exponentially when moving away from the bundle.

It will appear in the next sections that both the possibilities \mref{23a} and \mref{23b} will be met and will have
to be applied to solve the problem of the semiclassical quantization.

By the definition \mref{23} and by the recurrent equations
\mref{22f}-\mref{22g} and by the conditions \mref{23a}-\mref{23b} we have formulated the quantization procedure
which should determine skeletons, the SWF's $\Psi_{\bf B}^{as;\sigma}(x,y,\lambda)$ and the corresponding energy levels. In
the next two sections we shall apply this procedure to the simplest well known cases of billiards, i.e. to the circular
and the rectangular ones.

Let us note finally that if ${\bf B}\neq{\bf B}^A$ then energy levels corresponding to the skeleton ${\bf B}$ have
to be degenerate. This conclusion follows easily from the form of the quantization conditions \mref{22f}-\mref{22g} and
\mref{23a}-\mref{23b} showing that the complex conjugations of $\Psi_{\bf B}^{as;\sigma}(x,y,\lambda)$ satisfy also
these conditions with the same semiclassical energy $E$. The two corresponding solutions are of course
$\Psi_{\bf B}^{as;\pm}(x,y,\lambda)$.

\subsection{Finite and infinite bundle structures of skeletons. The last quantization condition}

\hspace{15pt}For a given billiards there can be skeletons with a finite number of bundles as well as with an infinite
one. The semiclassical quantization procedure described in the previous sections seems to be easily applied to
the finite bundle number skeletons. Namely in such a case following a trajectory starting from a bundle $B_k(u,l)$ we have
to approach the same bundle after a finite number of bounces. The corresponding semiclassical wave function propagated
by the skeleton has therefore to come back to its initial form achieving again the initial bundle. This condition
closes the process of quantization formulated in the previous sections. The respective conditions are of course the
following:
\be
(-\delta)^n\exp\ll(\sum_{B_k(u,l)\in {\bf B}}\phi_k(u,l)\r)=1\nn\\
\chi_k^{\sigma,cont}(D,s;u,l;\lambda)=\chi_k^\sigma(s;u,l;\lambda)
\label{23c}
\ee
where $n$ is a number of bounces and $D$ is the total distance passed by the billiards ball along the investigated
trajectory.

The cases of skeletons with an infinite number of bundles are much more difficult for investigations. Such skeletons
should be typical for chaotic billiards. According to its definition a bundles $B_k(u,l)$ can bifurcate after
the reflection by the billiards boundary into many different subbundles, i.e. parts of other bundles having
their beginnings also partly on the arc $BA_k(u,l)$. In fact a general behaviour of a skeleton in such chaotic cases
should not
differ essentially by its chaotic complexity from a chaotic trajectory reminding however rather a gigantic road-knot with
infinitely many viaducts spanning the billiards boundary on which the billiards ball moves. It is obvious that if they
exist their identification seems to be not an easy task.

Nevertheless the rule \mref{23c} can appear to be useful also even in such cases. This is because a ray beginning with
a bundle $B_k(u,l)$ can come back to it even arbitrarily close to its initial starting point on $A_k(u,l)$ (according to
the Poincare theorem) not repeating its way. But this is enough for writing the "last quantization condition" \mref{23c}
where the sum goes now over all bundles of the skeleton passed by the ray.

In the next sections we shall focus on the finite number cases of bundles in skeletons not avoiding however
billiards with chaotic motions such as the Bunimovich one. We shall discuss also in sec.8 a possibility of explaining
by the skeleton idea phenomena of scars. These are just phenomena which cannot certainly be described by skeletons with
finite structures.

\section{The circular billiards}

\hspace{15pt}In this section we would like to apply the procedure of constructing the SWF's on the classical
trajectories according to the last section to the circular billiards Fig.2 and to quantize
with these rules the energy of the circular billiard ball.

First we have to construct a skeleton {\bf B}. It follows from the considerations of App.B that any of it contains only
a single bundle $B(0,2\pi;\alpha)$ defined on the full circle $L(0,2\pi)$
rays of which make all the same angle $\alpha,\;0<\alpha\leq\fr\pi$, with the tangents to the circle. Its partner
${\bf B}^A$ contains also a single bundle $B(0,2\pi;\pi-\alpha)$ the rays of which make the angle $\pi-\alpha$ with
the tangents to the circle. Therefore in the case $\alpha=\fr\pi$ both the bundles coincide.

Motions along such rays
conserve of course the angular momentum $p\cos\alpha$ so that both the bundles respect the cylindrical symmetry of the
billiards.

\begin{figure}
\begin{center}
\psfig{figure=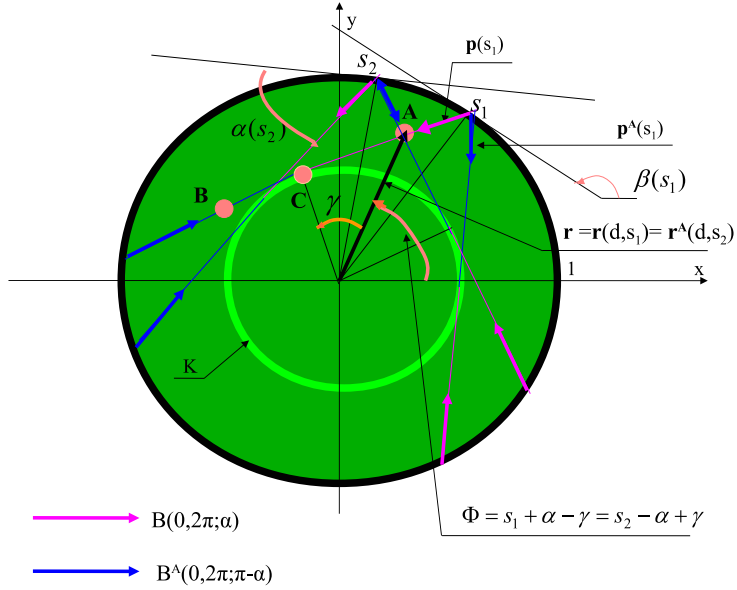,width=12cm}
\caption{The two bundle $B(0,2\pi;\alpha)$ and $B^A(0,2\pi;\pi-\alpha),\;0<\alpha,$ of the skeleton {\bf B} in the
circular billiards}
\end{center}
\end{figure}

For these two bundles the corresponding SWF's are the following (see Fig.2):
\be
\Psi^\pm(d,s,\lambda)=(d-\sin\alpha)^{-\fr}e^{\pm i\lambda p(d\pm s\cos\alpha)}\chi^\pm(d,s,\lambda)\nn\\
\Psi_A^\pm(d,s,\lambda)=(d-\sin\alpha)^{-\fr}e^{\pm i\lambda p (d\mp s\cos\alpha)}\chi_A^\pm(d,s,\lambda)
\label{B1}
\ee

The above SWF's are defined in the ring $0\leq d<\sin\alpha,\;0\leq s\leq 2\pi$ of the circular billiards.

Since both the bundles are defined in the ring mentioned the SWF's have to satisfy the following uniqueness conditions:
\be
\Psi^\pm(d,s,\lambda)=\Psi^\pm(d,s+2\pi,\lambda)\nn\\
\Psi_A^\pm(d,s,\lambda)=\Psi_A^\pm(d,s+2\pi,\lambda)
\label{B1a}
\ee
which lead to
\be
\lambda p\cos\alpha=-\lambda p\cos(\pi-\alpha)=m,\;\;\;\;\;\;\;\;m=0,\pm 1,\pm 2,...
\label{B2}
\ee
and means the angular momentum quantization together with
\be
\chi^\pm(d,s+2\pi,\lambda)=\chi^\pm(d,s,\lambda)\nn\\
\chi_A^\pm(d,s+2\pi,\lambda)=\chi_A^\pm(d,s,\lambda)
\label{B2a}
\ee

We can demand from the SWF's $\Psi^\pm(d,s,\lambda)$ and $\Psi_A^\pm(d,s,\lambda)$ to be eigenfunctions
$\Psi_{\pm m}(d,s,\lambda)$ ($=\Psi^\pm(d,s,\lambda)$) and $\Psi_{A;\pm m}(d,s,\lambda)$ ($=\Psi_A^\mp(d,s,\lambda)$) of the
respective angular momenta $\pm m\hbar$, $m=0,1,2,...$, where $m=\lambda p\cos\alpha$. Then it follows easily from the
forms of the SWF's considered
that for the corresponding $\chi_{\pm m}(d,s,\lambda)$ and $\chi_{A;\pm m}(d,s,\lambda)$ we have to have:
\be
\frac{\p\chi_{\pm m}(d,s,\lambda)}{\p s}=
\frac{\p\chi_{A;\pm m}(d,s,\lambda)}{\p s}=0
\label{B2b}
\ee
i.e. both $\chi_{\pm m}(d,s,\lambda)$ and $\chi_{A;\pm m}(d,s,\lambda)$ are independent of $s$ and
therefore they are constant on the circle boundary, i.e. for $d=0$.

Putting therefore $\chi_m^\pm(0,s,\lambda)\equiv 1$ and assuming further the following boundary conditions:
\be
\chi_{\pm m}(0,s,\lambda)=-\chi_{A;\pm m}(0,s,\lambda)\equiv 1
\label{B2c}
\ee
we get for the appropriate $\Psi_{\pm m}^{as}(x,y,\lambda)$ vanishing on the circle boundary and being the angular
momentum eigenfunction:
\be
\Psi_{\pm m}^{as}(x,y,\lambda)=\Psi_{\pm m}(d,s_1,\lambda)+\Psi_{A;\pm m}(d,s_2,\lambda)\nn\\
m=0,1,2,...\nn\\
(x,y)=(\cos s_1,\sin s_1)+d(-\sin(\alpha+s_1),\cos(\alpha+s_1))=\nn\\
(\cos s_2,\sin s_2)-d(-\sin(-\alpha+s_2),\cos(-\alpha+s_2))
\label{B2d}
\ee

Making further the identification
\be
\Psi_{A;\pm m}(d,s,\lambda)=\Psi_{\pm m}^{cont}(2\sin\alpha-d,s-2\alpha,\lambda),\;\;\;\;\sin\alpha> d\geq 0
\label{B4}
\ee
where $\Psi_{\pm m}^{cont}(d,s,\lambda)$ denote $\Psi_{\pm m}(d,s,\lambda)$ continued by the caustic $r=\cos\alpha$ we
get for $\Psi_{\pm m}^{as}(x,y,\lambda)\equiv\Psi_{\pm m}^{as}(r\cos\Phi,r\sin\Phi,\lambda)$ (see Fig.2):
\be
\Psi_{\pm m}^{as}(r\cos\Phi,r\sin\Phi,\lambda)=\Psi_{\pm m}(d,s,\lambda)+\Psi_{A;\pm m}(d,2\Phi-s,\lambda)=\nn\\
\Psi_{\pm m}(d,s,\lambda)+\Psi_{\pm m}^{cont}(2\sin\alpha-d,2\Phi-s-2\alpha,\lambda)=\nn\\
(\sin\alpha-d)^{-\fr}\ll(-ie^{\pm i\lambda pd\pm ims}\chi_{\pm m}(d,s,\lambda)\r.\pm\nn\\
\ll.e^{\mp i\lambda p(d-2\sin\alpha)\pm im(2\Phi-s-2\alpha)}\chi_{\pm m}^{cont}(2\sin\alpha-d,2\Phi-s-2\alpha,\lambda)\r)=\nn\\
m=0,1,2,...\nn\\
d=\sin\alpha-\sqrt{r^2-\cos^2\alpha}\nn\\
\sin(\Phi-s)=\frac{\sin\alpha}{r}(\sin\alpha-\sqrt{r^2-\cos^2\alpha})\nn\\
\sin\alpha>d\geq 0
\label{B3}
\ee

The identification conditions \mref{B4} give:
\be
\chi_{A;\pm m}(d,s,\lambda)=e^{\pm 2i(\lambda p\sin\alpha-m\alpha-\frac{\pi}{4})}
\chi_{\pm m}^{cont}(2\sin\alpha-d,s-2\alpha,\lambda)\nn\\
\sin\alpha> d\geq 0
\label{B5}
\ee

Applying now \mref{23c} and \mref{B2c} we get:
\be
1=e^{\pm 2i(\lambda p\sin\alpha-m\alpha+\frac{\pi}{4})}\nn\\
\chi_{\pm m}^{cont}(2\sin\alpha,s-2\alpha,\lambda)=\chi_{\pm m}(0,s,\lambda)\equiv 1
\label{B5a}
\ee
so that
\be
\int_0^{2\sin\alpha}\ll(J^{\fr}\triangle\ll(J^{-\fr}{\chi}_{\pm m;k}(a,s)\r)+
2\sum_{l=0}^kE_{k-l+1}{\chi}_{\pm m;l}(a,s)\r)da=0\nn\\
k=0,1,...
\label{B6}
\ee
with
\be
\chi_{\pm m;0}(d,s)\equiv 1\nn\\
\chi_{\pm m;k+1}(d,s)=\pm\frac{i}{2p}\int_0^d\ll(J^{\fr}\triangle\ll(J^{-\fr}\chi_{\pm m;k}(a,s)\r)+
2\sum_{l=0}^kE_{k-l+1}\chi_{\pm m;l}(a,s)\r)da\nn\\
k=0,1,2,...
\label{B6a}
\ee

The integrations in the formulae \mref{B6}-\mref{B6a}  go above the singular point $d=\sin\alpha$ in the
$d$-plane for the plus sign and below this point for the minus one.

A simple conclusion from the quantization conditions \mref{B6}-\mref{B6a} is that:
\be
\ll(\chi_{+m;k}(d,s)\r)^*=\chi_{-m;k}(d,s)\nn\\
k=0,1,2,...
\label{B7}
\ee
so that
\be
\ll(\Psi_{+m}^{as}(x,y,\lambda)\r)^*=\Psi_{-m}^{as}(x,y,\lambda)
\label{B7a}
\ee
i.e. the semiclassical energies $E$ in the circular billiards are degenerate with respect to the sign of the angular
momentum.

Of course the above conclusions are well known as exact for the circular billiards.

Therefore in the equations \mref{B6}-\mref{B6a} we can choose only the plus sign in the respective calculations to get:
\be
\lambda\sqrt{2E_{0;mr}-\frac{m^2}{\lambda^2}}-m\arccos\frac{m}{\lambda\sqrt{2E_0}}=r\pi-\frac{\pi}{4}\nn\\
\cos\alpha_{mr}=\frac{m}{\lambda\sqrt{2E_{0;mr}}}\nn\\
m=0,1,2,...,\;\;\;\;r=1,2,...
\label{B7b}
\ee
for the "classical" energy $E_0$ and the angle $\alpha$.

It follows from \mref{B7b} that
\be
E_{0;0r}=\frac{\ll(r-\frac{1}{4}\r)^2\pi^2}{2\lambda^2}\nn\\
\cos\alpha_{0r}=0\nn\\
E_{0;mr}=\frac{m^2}{2\lambda^2F^2(\frac{r-\frac{1}{4}}{m}\pi)}\nn\\
\cos\alpha_{mr}=F(\frac{r-\frac{1}{4}}{m}\pi)\nn\\
m=1,2,...,\;\;\;\;r=1,2,...
\label{B7c}
\ee
where $F^{-1}(x)=(\sqrt{1-x^2}-x\arccos x)x^{-1},\;0<x\leq 1$.

The next to zero order terms of the corresponding semiclassical expansions can be obtained using the results of App.D
from which it follows (the formulae \mref{A3}-\mref{A4}) that the operator $D(t,s)=J^{\fr}\cdot\triangle\cdot J^{-\fr}$
in \mref{B8} is bilinear in $\frac{\p}{\p t}$ and $\frac{\p}{\p s}$ with coefficients completely independent of $s$ and
it acts in \mref{B8} also on the $s$-independent quantities. In such cases its action is reduced to (for $t\to x=pt$):
\be
\ll[J^{\fr}\cdot\triangle\cdot J^{-\fr}\r]_{red}(x,\frac{\p}{\p x})=\nn\\
\ll(\frac{\cos^2\alpha}{J^2}+1\r)\frac{\p^2}{\p x^2}+
\frac{2\cos^2\alpha}{J^3}\frac{\p}{\p x}+\frac{5}{4}\frac{\cos^2\alpha}{J^4}-
\frac{1}{4}\frac{1}{J^2}
\label{B9}
\ee
where $J=\sin\alpha-x$.

Taking this into account in the second of the eqs. \mref{B6} and putting there $k=0$ we get:
\be
E_{1;mr}=-\frac{1}{4\sin\alpha_{mr}}\int_0^{2\sin\alpha_{mr}}\ll(J^{\fr}\cdot\triangle\cdot J^{-\fr}\r)_{red}\cdot 1dx=
\frac{1}{8\sin\alpha_{mr}}\ll(\frac{5}{6}\cot^2\alpha_{mr}-1\r)
\label{B8}
\ee
where the integration in the $x$-plane went over upper half-circle with the center at $x=\sin\alpha_{mr}$ in the
clockwise direction.

Similarly for $\chi_{mr;1}(d)$ we get:
\be
\chi_{mr;1}(d)=\frac{i}{2p}\int_0^d\ll(\frac{5}{4}\frac{\cos^2\alpha_{mr}}{J_{mr}^4}-
\frac{1}{4}\frac{1}{J_{mr}^2}\r)dx+\frac{i}{p}E_{1;mr}d=\nn\\
-\frac{i}{2p}\ll(\frac{5}{12}\frac{\cos^2\alpha_{mr}}{J_{mr}^3}-\frac{5}{12}\frac{\cos^2\alpha_{mr}}{\sin^3\alpha_{mr}}-
\frac{1}{4}\frac{1}{J_{mr}}+\frac{1}{4}\frac{1}{\sin\alpha_{mr}}\r)+\frac{i}{p}E_{1;mr}d
\label{B10a}
\ee
where $J_{mr}=\sin\alpha_{mr}-d$.

The higher order terms of the semiclassical expansions for $E_{mr}(\lambda)$ and $\chi_{mr}(d,\lambda)$ can be obtain analogously
using the recurrent equations \mref{B6}-\mref{B6a}.

It is to be noticed that the above calculations form a new algorithm for the semiclassical approximation method.

\section{The rectangular billiards}

\hskip+2em Consider now the rectangular billiards shown in Fig.3. This billiards is the canonical example of the energy
quantization problem because of its easiness to be solved by the variable separation method. According to
Fig.3 the well known solution to the problem is given by the following two equations:
\be
\sin(\lambda p_xa)=0\nn\\
\sin(\lambda p_yb)=0
\label{A0}
\ee
with $p_x^2+p_y^2=p^2=2E$ as the result of the following form of the energy eigenfunction:
\be
\Psi(x,y)=C\sin(\lambda p_xx)\sin(\lambda p_yy)=\nn\\
-\frac{C}{4}\ll(e^{\lambda p_xx+\lambda p_yy}+
e^{-\lambda p_xx-\lambda p_yy}-e^{\lambda p_xx-\lambda p_yy}-e^{-\lambda p_xx+\lambda p_yy}\r)
\label{A0a}
\ee

Of course one can always put $p_x=p\cos\alpha,p_y=p\sin\alpha$ where $\alpha,\;0<\alpha<\fr\pi$, is the
angle the momentum $p$ is inclined to the $x$-axis.

Let us note that the cases $\alpha=0,\fr\pi$ are excluded by the solutions \mref{A0a}.
\begin{figure}
\begin{center}
\psfig{figure=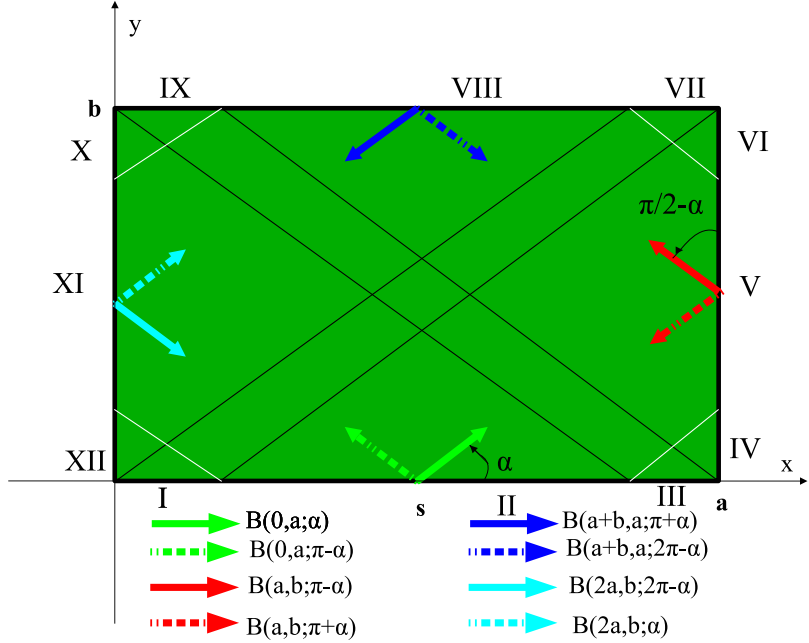,width=12cm}
\caption{The rectangular billiards with eight ray bundles}
\end{center}
\end{figure}

\subsection{The cases $0<\alpha<\fr\pi$}

\hskip+2em Let us consider this rectangular billiard example by the method described in the previous sections. Since the absolute
values of the momentum components $p_x,p_y$ are the integrals of the classical motion inside the billiards respecting
elastic law of bouncing then all bundles which should be taken into account are defined by a single angle
$\alpha,\;0<\alpha<\fr\pi$, which are made by the rays of the bundle $B_1=B(0,a;\alpha)$ with the $x$-axis, see Fig.3.

Choosing the case of the angle $\alpha$ shown in Fig.3 the remaining seven bundles of the skeleton {\bf B} shown in Fig.3  are
$B_2=B(0,a;\pi-\alpha),\;B_3=B(a,b;\pi-\alpha),\;B_4=B(a,b;\pi+\alpha),\;B_5=B(a+b,a;\pi+\alpha),\;B_6=B(a+b,a;2\pi-\alpha),\;
B_7=B(2a+b,b;2\pi-\alpha),\;B_8=B(2a+b,b;\alpha)$, i.e. the parameter $s$ introduced in sec.3 is counted anticlockwise
starting from the
point $(0,0)$ of Fig.3 (and having negative value if measured clockwise). The bundles $B_{2k-1},\;B_{2k},\;k=1,...,4,$
are defined on the respective sides $L_k,\;k=1,...,4,$ of
the billiards, i.e. on $L_1=L(0,a),\;L_2=L(a,b),\;L_3=L(a+b,a),\;L_4=L(2a+b,b)$.

The skeleton ${\bf B}^T$ coincides exactly with {\bf B} in the case of the rectangular billiards.

Let us note that a number of bundles in the skeletons is obviously independent of a choice of $\alpha$, i.e. it is
always equal to eight.

Note also that unlike the circular billiard bundles the rectangular bundles do not have
their caustics inside the billiards, but rather in the infinities.

The semiclassical wave function $\Psi^{as}(x,y)$ constructed according the rules discussed earlier should be the sum
at most of the eight SWF's $\Psi_i^+(d,s,\lambda),\;i=1,...,8$, which are constructed on the respective bundles $B_i,\;i=1,...,8$,
of the skeleton {\bf B}. But as it follows from Fig.4 only four of them interfere in each point of the rectangular billiards, i.e. we have:
\be
\Psi^{as}(x,y)=\sum_{i=1}^8\Psi_i^+(d_i,s_i,\lambda)=\sum_{i=1}^4\Psi_{k_i}^+(d_i,s_i,\lambda)\nn\\
x=x_0(s_i)+d_i\cos\alpha(s_i)\nn\\
y=y_0(s_i)+d_i\sin\alpha(s_i)\nn\\
i=1,...,4
\label{A1}
\ee
where $\alpha(s_i)$ are the angles the rays in Fig.4 make with the $x$-axis and $(x_0(s_i),y_0(s_i))$ are the points of
the rectangular boundary from which the rays start. The parameters $s_i$ as previously measure distances of the boundary
points $(x_0(s_i),y_0(s_i))$
from the point $(0,0)$ in anticlockwise direction along the boundary.
The corresponding Jamaicans are $J(t,s_i)=\sqrt{-\sin\alpha(s_i)}$, i.e. are constant but discontinues. Therefore they
will be included into the $\chi$-factors contained in the SWF's.
\begin{figure}
\begin{center}
\psfig{figure=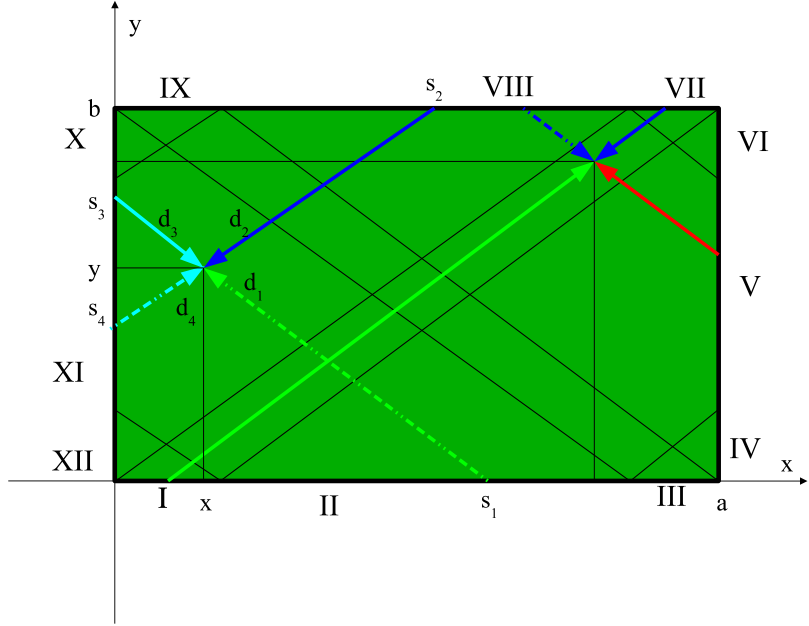,width=12cm}
\caption{Four solutions $\Psi_i^+(d_i,s_i),\;i=1,...,4$, meeting at the point ($x,y$)}
\end{center}
\end{figure}

Let us note that $\Psi^{as}(x,y)$ can be discontinue each time the point $(x,y)$ crosses the thin black lines in
Fig.3. being the boundary of different bundles.
Let us enumerate these lines by $l_i,\;i=1,...,4,$ starting from the line emerging from the point $(0,0)$ and continuing
anticlockwise.

It is then easy to see that to ensure the continuity of $\Psi^{as}(x,y)$ in each point in the billiards it is
enough to identify the respective $\chi$-coefficients on the respective lines defining them elsewhere in the billiards
as continuous functions of $s$ and $d$. We get such a continuity putting:
\be
\chi_1^\pm(d,0,\lambda)=\chi_8^\pm(d,0,\lambda)\nn\\
\chi_3^\pm(d,a,\lambda)=\chi_2^\pm(d,a,\lambda) \nn\\
\chi_5^\pm(d,a+b,\lambda)=\chi_4^\pm(d,a+b,\lambda)\nn\\
\chi_7^\pm(d,2a+b,\lambda)=\chi_6^\pm(d,2a+b,\lambda)
\label{A1a}
\ee

Of course $\Psi^{as}(x,y)$ vanishes on the rectangular boundary, i.e. we have to have:
\be
\Psi^{as}(0,y)=\Psi^{as}(a,y)=\Psi^{as}(x,0)=\Psi^{as}(x,b)=0\nn\\
0<x<a,\;\;\;0<y<b
\label{A2}
\ee

The conditions \mref{A2} expressed in terms of the SWF's $\Psi_i^\pm(d,s,\lambda),\;i=1,...,8,$ take the following detailed forms:
\be
\Psi_{2k-1}^+(0,s,\lambda)+\Psi_{2k}^-(0,s,\lambda)=0\nn\\
\Psi_{2k}^+(0,s,\lambda)+\Psi_{2k-1}^-(0,s,\lambda)=0\nn\\
s\in L_k,\;\;\;\;\;\;\;k=1,...,4
\label{A2a}
\ee

When identifying $\Psi_i^-(0,s,\lambda),\;i=1,...,8$, with the respective $\Psi_j^+(d_j(s),f_j(s),\lambda),\;j=1,...,8$, it
should be noticed that for each considered trajectory family there is no caustic inside the rectangular billiards
so that each integral $\int_Kp_xdx+p_ydy$ along the close loop $K$ lying
inside the billiards or on its boundary has to vanish. As a consequence of this the identifications we are talking about
are then reduced just to the identifications of the corresponding $\chi$-factors, i.e the possible phase factors are
factored out. Therefore we get successively for the conditions \mref{A2a} to be satisfied:
\begin{enumerate}
\item $s>0,\;{\bf r}_0(s)\in I,\;-s\tan\alpha\in XII,\;2a+b-b\cot\alpha-{\bf r}_0(s)\in VII$
\be
\chi_1^-(0,s,\lambda)=\chi_4^+\ll(\frac{b}{\sin\alpha},2a+b-b\cot\alpha-s,\lambda\r)\nn\\
\chi_2^-(0,s,\lambda)=\chi_7^+\ll(\frac{s}{\cos\alpha},-s\tan\alpha,\lambda\r)
\label{A2f}
\ee
and hence and from \mref{A2a}
\be
\chi_1(0,s,\lambda)+\chi_7\ll(\frac{s}{\cos\alpha},-s\tan\alpha,\lambda\r)=0\nn\\
\chi_2(0,s,\lambda)+\chi_4\ll(\frac{b}{\sin\alpha},2a+b-b\cot\alpha-s,\lambda\r)=0
\label{A3}
\ee

In the last two equations the plus signs at $\chi_j^+,\;j=1,...,8$, have been dropped and this convention will be
kept in further equations if not leading to misunderstandings.

\item $s>0,\;{\bf r}_0(s)\in II,\;-s\tan\alpha\in X+XI,\;a+(a-s)\tan\alpha\in V+VI)$
\be
\chi_1^-(0,s,\lambda)=\chi_4^+\ll(\frac{a-s}{\cos\alpha},a+(a-s)\tan\alpha,\lambda\r)\nn\\
\chi_2^-(0,s,\lambda)=\chi_7^+\ll(\frac{s}{\cos\alpha},-s\tan\alpha,\lambda\r)
\label{A3a}
\ee
and hence and from \mref{A2a}
\be
\chi_1(0,s,\lambda)+\chi_7\ll(\frac{s}{\cos\alpha},-s\tan\alpha,\lambda\r)=0\nn\\
\chi_2(0,s,\lambda)+\chi_4\ll(\frac{a-s}{\cos\alpha},a+(a-s)\tan\alpha,\lambda\r)=0
\label{A4}
\ee
\item ${\bf r}_0(s)\in III,\;s>0,\;(-b-s+b\cot\alpha\in IX,\;a+(a-s)\tan\alpha\in IV$
\be
\chi_1^-(0,s,\lambda)=\chi_4^+\ll(\frac{a-s}{\cos\alpha},a+(a-s)\tan\alpha,\lambda\r)\nn\\
\chi_2^-(0,s,\lambda)=\chi_6^+\ll(\frac{b}{\sin\alpha},-b-s+b\cot\alpha,\lambda\r)
\label{A4a}
\ee
and hence
\be
\chi_1(0,s,\lambda)+\chi_6\ll(\frac{b}{\sin\alpha},-b-s+b\cot\alpha,\lambda\r)=0\nn\\
\chi_2(0,s,\lambda)+\chi_4\ll(\frac{a-s}{\cos\alpha},a+(a-s)\tan\alpha,\lambda\r)=0
\label{A5}
\ee
\item $s>0,\;{\bf r}_0(s)\in IV+V+VI,\;a-(s-a)\cot\alpha\in II+III,\\a+b+(a+b-s)\cot\alpha\in VII+VIII$
\be
\chi_3^-(0,s,\lambda)=\chi_6^+\ll(\frac{a+b-s}{\sin\alpha},a+b+(a+b-s)\cot\alpha,\lambda\r)\nn\\
\chi_4^-(0,s,\lambda)=\chi_1^+\ll(\frac{s-a}{\sin\alpha},a-(s-a)\cot\alpha,\lambda\r)
\label{A5a}
\ee
and hence
\be
\chi_3(0,s)e^{-i\lambda pa\cos\alpha}+\chi_1\ll(\frac{s-a}{\sin\alpha},a-(s-a)\cot\alpha,\lambda\r)e^{i\lambda pa\cos\alpha}=0\nn\\
\chi_4(0,s,\lambda)e^{-i\lambda pa\cos\alpha}+
\chi_6\ll(\frac{a+b-s}{\sin\alpha},a+b+(a+b-s)\cot\alpha,\lambda\r)e^{i\lambda pa\cos\alpha}=0
\label{A6}
\ee
\item $s>0,\;{\bf r}_0(s)\in VII,\;2a+b-b\cot\alpha-{\bf r}_0(s)\in I,\;a+b-(s-a-b)\tan\alpha\in VI$
\be
\chi_5^-(0,s,\lambda)=\chi_1^+\ll(\frac{b}{\sin\alpha},2a+b-b\cot\alpha-s,\lambda\r)\nn\\
\chi_6^-(0,s,\lambda)=\chi_3^+\ll(\frac{s-a-b}{\cos\alpha},a+b-(s-a-b)\tan\alpha,\lambda\r)
\label{A6a}
\ee
and hence
\be
\chi_6(0,s,\lambda)e^{-i\lambda pb\sin\alpha}+
\chi_1\ll(\frac{b}{\sin\alpha},2a+b-b\cot\alpha-s,\lambda\r)e^{i\lambda pb\sin\alpha}=0\nn\\
\chi_5(0,s,\lambda)e^{-i\lambda pb\sin\alpha}+
\chi_3\ll(\frac{s-a-b}{\cos\alpha},a+b-(s-a-b)\tan\alpha,\lambda\r)e^{i\lambda pb\sin\alpha}=0
\label{A7}
\ee
\item $s<0,\;{\bf r}_0(s)\in VIII,\;-b-(b+s)\tan\alpha\in X+XI+XII,\\a+b-(a+b+s)\tan\alpha\in IV+V$
\be
\chi_3^-(0,s)=\chi_1^+\ll(\frac{-b-s}{\cos\alpha},-b-(b+s)\tan\alpha,\lambda\r)\nn\\
\chi_4^-(0,s)=\chi_2^+\ll(\frac{a+b+s}{\cos\alpha},a+b-(a+b+s)\tan\alpha,\lambda\r)
\label{A7a}
\ee
and hence
\be
\chi_4(0,s,\lambda)e^{-i\lambda pb\sin\alpha}+
\chi_1\ll(\frac{-b-s}{\cos\alpha},-b-(b+s)\tan\alpha,\lambda\r)e^{+i\lambda pb\sin\alpha}=0\nn\\
\chi_3(0,s,\lambda)e^{-i\lambda pb\sin\alpha}+
\chi_2\ll(\frac{a+b+s}{\cos\alpha},a+b-(a+b+s)\tan\alpha,\lambda\r)e^{i\lambda pb\sin\alpha}=0
\label{A8}
\ee
\item $s<0,\;{\bf r}_0(s)\in IX,\;-b-(b+s)\tan\alpha\in X,\;b\cot\alpha-b-{\bf r}_0(s)\in III$
\be
\chi_5^-(0,s)=\chi_8^+\ll(\frac{-b-s}{\cos\alpha},-b-(b+s)\tan\alpha,\lambda\r)\nn\\
\chi_6^-(0,s)=\chi_2^+\ll(\frac{b}{\sin\alpha},b\cot\alpha-b-s,\lambda\r)
\label{A8a}
\ee
and hence
\be
\chi_6(0,s,\lambda)e^{-i\lambda pb\sin\alpha}+
\chi_8\ll(\frac{-b-s}{\cos\alpha},-b-(b+s)\tan\alpha,\lambda\r)e^{+i\lambda pb\sin\alpha}=0\nn\\
\chi_5(0,s,\lambda)e^{-i\lambda pb\sin\alpha}+
\chi_2\ll(\frac{b}{\sin\alpha},b\cot\alpha-b-s,\lambda\r)e^{i\lambda pb\sin\alpha}=0
\label{A9}
\ee
\item $s<0,\;{\bf r}_0(s)\in X+XI+XII,\;-s\cot\alpha\in I+II,\\-b-(s+b)\cot\alpha\in VIII+IX$
\be
\chi_8^-(0,s,\lambda)=\chi_5^+\ll(\frac{b+s}{\sin\alpha},-b-(s+b)\cot\alpha,\lambda\r)\nn\\
\chi_7^-(0,s,\lambda)=\chi_2^+\ll(\frac{-s}{\sin\alpha},-s\cot\alpha,\lambda\r)
\label{A9a}
\ee
and hence
\be
\chi_7(0,s,\lambda)+\chi_2\ll(\frac{-s}{\sin\alpha},-s\cot\alpha,\lambda\r)=0\nn\\
\chi_8(0,s,\lambda)+
\chi_3\ll(\frac{b+s}{\sin\alpha},-b-(s+b)\cot\alpha,\lambda\r)=0
\label{A10}
\ee
\end{enumerate}

Solving the last set of the quantization conditions we should get the solutions for the energy levels and for the
corresponding $\Psi^{as}(x,y)$.

However the above conditions do not provide us with the solutions for $\chi_i(0,s,\lambda),\;i=1,...,8$,
in some algebraic forms. Rather they show how a solution defined on some bundle propagates along its rays
to be transformed by bouncing off the rectangular boundary into another bundle and finally after a finite number
of such bouncings achieving its mother bundle and repeating this process infinitely.

\begin{figure}
\begin{center}
\psfig{figure=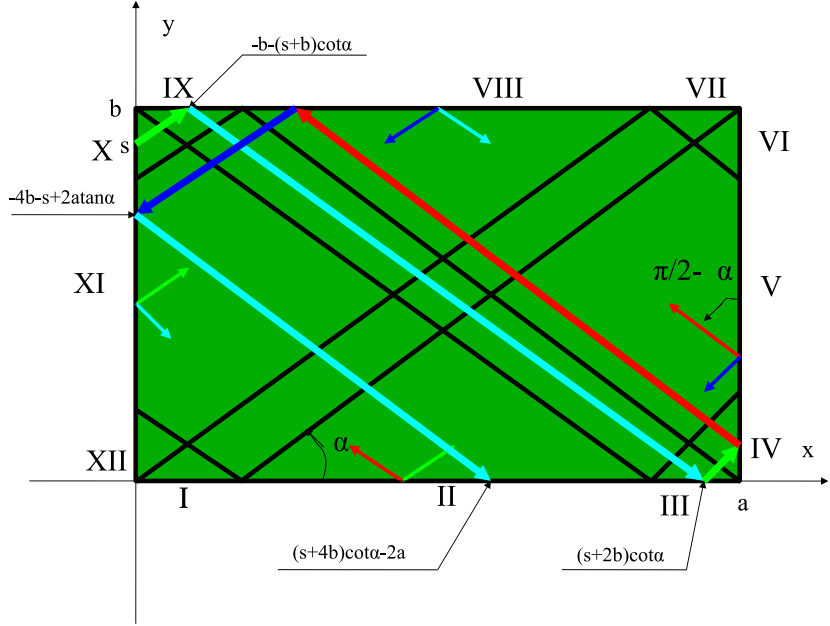,width=12cm}
\caption{The solution $\Psi_8^+(t,s)$ being carried by the successive bundles
$B_8\to B_6\to B_1\to B_3\to B_5\to B_7\to B_1$}
\end{center}
\end{figure}

To get a knowledge what really happens by such bouncings
consider for example the green color ray along which the solution $\Psi_8^+(t,s)$ propagates, shown
in Fig.5 emerging from the point $s,\;{\bf r}_0(s)\in X$, of the rectangular boundary. This particular ray bounces into
the blue-light one and next again into the green one. Taking into account first the conditions \mref{A7} and
next \mref{A5} we get:
\be
\chi_1(0,(s+2b)\cot\alpha,\lambda)=-\chi_6\ll(\frac{b}{\sin\alpha},-b-(s+b)\cot\alpha,\lambda\r)=\nn\\
\chi_8^{cont}\ll(\frac{2b+s}{\sin\alpha},s,\lambda\r)e^{+2i\lambda pb\sin\alpha},\;\;\;\;\;{\bf r}_0(s)\in X
\label{A11}
\ee
and continuing the bounces to achieve the green ray again we get:
\be
\chi_1(0,(s+4b)\cot\alpha-2a,\lambda)=
\chi_8^{cont}(D(s),s,\lambda)e^{4i\lambda pb\sin\alpha+2i\lambda pa\cos\alpha},\;\;\;\;\;{\bf r}_0(s)\in X
\label{A12}
\ee
where $D(s)$ is the total distance the green ray has passed from the point $s$ to the point
$(s+4b)\cot\alpha-2a$ bouncing multiply from the boundary.

But taking into account the first of the identification \mref{A1a} and the identification \mref{23c} we have to have:
\be
\chi_1(0,(s+2b)\cot\alpha,\lambda)=\chi_8^{cont}\ll(\frac{2b+s}{\sin\alpha},s,\lambda\r)\nn\\
\chi_1(0,(s+4b)\cot\alpha-2a,\lambda)=\chi_8^{cont}(D(s),s,\lambda)\nn\\
{\bf r}_0(s)\in X
\label{A13}
\ee
so that
\be
e^{+2i\lambda pb\sin\alpha}=1\nn\\
e^{4i\lambda pb\sin\alpha+2i\lambda pa\cos\alpha}=1
\label{A14}
\ee
and consequently:
\be
\lambda pb\sin\alpha=m\pi\nn\\
\lambda pa\cos\alpha=n\pi\nn\\
m,n=1,2,...
\label{A14a}
\ee

Taking the limit $\lambda\to\infty$ in \mref{A13} we get:
\be
\chi_{1,0}((s+2b)\cot\alpha)=\chi_{1,0}((s+4b)\cot\alpha-2a)=\chi_{8,0}(s)
\label{A15}
\ee

If $\tan\alpha=\frac{m}{n}\frac{a}{b}$ is not rational none of the arrival points can be repeated, i.e. the
corresponding trajectory is not closed (periodic), and the (infinite) set of such points is dense on the side $L_1$ of
the billiards. Therefore the equations \mref{A15} show that $\chi_{1,0}(s)$ has to be not only a constant of motion but
also $s$-independent since this equation can be written in infinitely many points of $L_1$ densely distributed on it.

The last conclusion is valid also of course for the remaining $\chi_{k,0}(s),\;k=1,...,8,$ and as it can be easily
concluded from \mref{12z} and App.D all the coefficients $\chi_k(d,s,\lambda),\;k=1,...,8,$ are then also constant.
Putting therefore $\chi_1(d,s,\lambda)\equiv 1$ we get by \mref{A1a} and \mref{A2a}:
\be
\chi_k(d,s,\lambda)\equiv 1,\;\;\;\;\;\;\;k=1,4,5,8\nn\\
\chi_k(d,s,\lambda)\equiv -1,\;\;\;\;\;\;\;k=2,3,6,7
\label{A16}
\ee

Choosing therefore the point $(x,y)$ of Fig.4 for the SWF \mref{A1} we get:
\be
\Psi^{as}(x,y)=\Psi_2^+(d_1,s_1)+\Psi_5^+(d_2,s_2)+\Psi_7^+(d_3,s_3)+\Psi_8^+(d_4,s_4)=\nn\\
\sum_{l=1}^4(-1)^le^{i\lambda p^2t_l+i\lambda pf_l(s_l)}
\label{A17}
\ee
where $t_l$ and $f_l(s_l),\;l=1,...,4$ should be calculated from the relations (see Fig.5):
\be
x=x_0(s_l)+pt_l\cos\alpha(s_l)\nn\\
y=y_0(s_l)+pt_l\sin\alpha(s_l)\nn\\
\alpha(s_l)=\alpha,\;\fr\pi-\alpha
\label{A18}
\ee

However making use of the independence of the phase integral $\int_{(0,0)}^{(x,y)}p_xdx+p_ydy$ of the integration
paths we get for the particular terms in the sum in \mref{A17}:
\be
e^{i\lambda p^2t_1+i\lambda pf_1(s_1)}=e^{i\lambda(-p_xx+p_yy)}\nn\\
e^{i\lambda p^2t_2+i\lambda pf_2(s_2)}=e^{i\lambda(-p_xx-p_yy)}\nn\\
e^{i\lambda p^2t_3+i\lambda pf_3(s_3)}=e^{i\lambda(p_xx-p_yy)}\nn\\
e^{i\lambda p^2t_4+i\lambda pf_4(s_4)}=e^{i\lambda(p_xx+p_yy)}
\label{A19}
\ee
and finally:
\be
\Psi^{as}(x,y)=-e^{i\lambda(-p_xx+p_yy)}+e^{i\lambda(-p_xx-p_yy)}-e^{i\lambda(p_xx-p_yy)}+e^{i\lambda(p_xx+p_yy)}=\nn\\
-4\sin(p_xx)\sin(p_yy)
\label{A20}
\ee
reproducing in this way the exact result \mref{A0a}.

The cases when $\tan\alpha=\frac{m}{n}\frac{a}{b}$ is rational are possible only if $\frac{a}{b}$ is rational.
For maintaining the results for the irrational $\frac{a}{b}$ one can argue relying on a continuity of all the
investigated quantities considered as functions of $a,b$, since
each point $(a,b)$ with rational value of $\frac{a}{b}$ is densely surrounded by the ones for which $\frac{a}{b}$ is
irrational.

\subsection{The cases $\alpha=0,\;\fr\pi$ - the bouncing mode skeletons}

\hskip+2em It is surprising that the cases $\alpha=0,\;\fr\pi$ are not allowed by the representation \mref{A0a} of the
SWF leading to the totally vanishing solutions while there are still the bundle skeletons with these angles on which
non vanishing identically SWF's can be constructed. Therefore we should get some new knowledge about the semiclasscical
method developed here considering these cases known as the bouncing modes.

Consider the case $\alpha=\fr\pi$. The second case will be analogous.

There are two bundles in this case which the skeleton ${\bf B}$ is consisting of. The one $B_1$ with its rays directed up and
starting from the side $L_1$ and the second $B_3$ with rays directed down starting from the side $L_3$ (Fig.6). The skeleton
${\bf B}^A$ is identical with ${\bf B}$.
\begin{figure}
\begin{center}
\psfig{figure=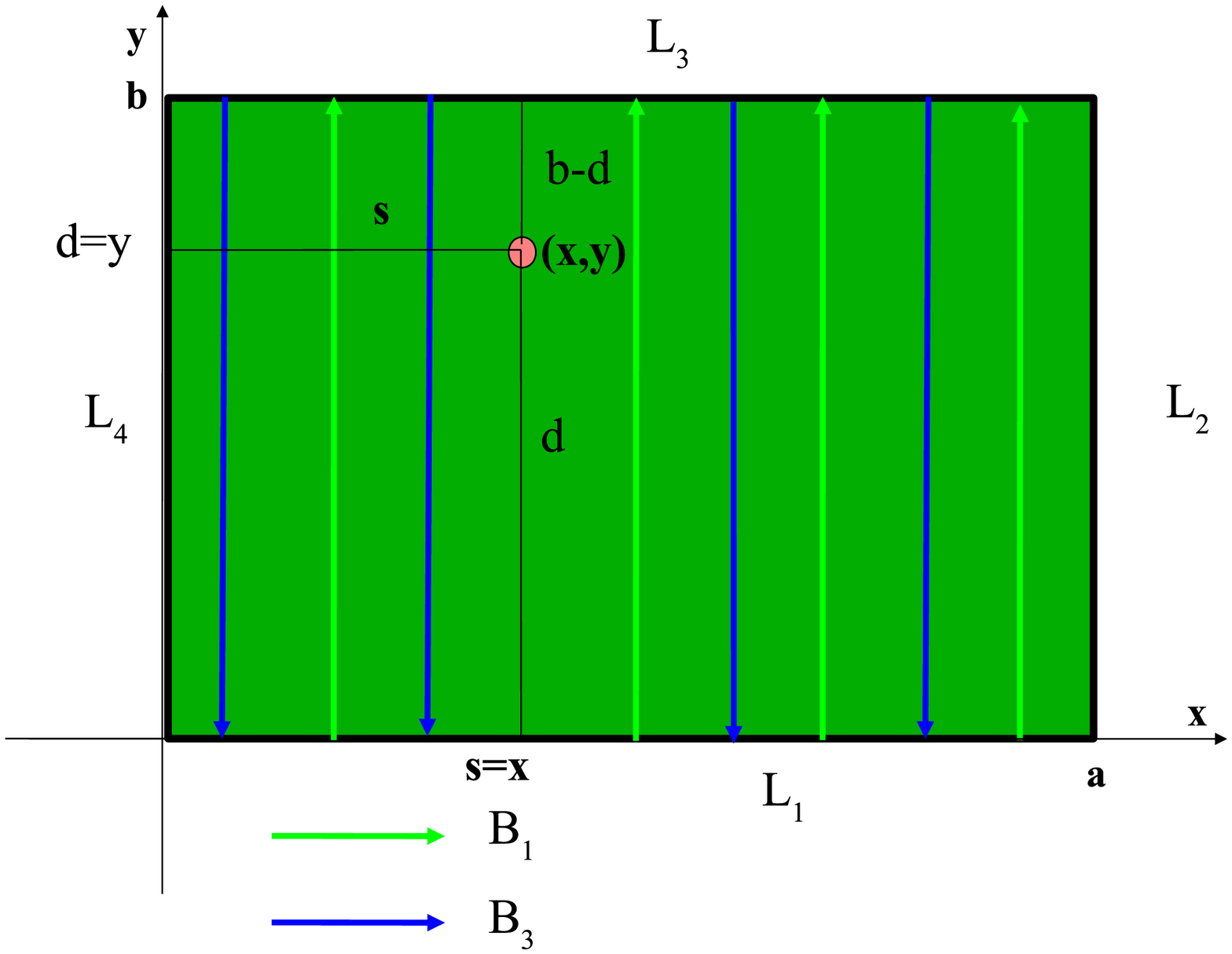,width=12cm}
\caption{The two bouncing mode bundles $B_1$ and $B_3$ of the vertical skeleton in the rectangular billiards}
\end{center}
\end{figure}

For these particular cases of bundles rays for both the bundles will be positioned by the same parameter $s$ measuring
a distance of a ray from the $y$-axis along the corresponding sides $L_1$ and $L_3$. Therefore for the corresponding SWF's
we get:
\be
\Psi_1^\pm(d,s,\lambda)=e^{\pm i\lambda pd}\chi_1^\pm(d,s,\lambda)\nn\\
\Psi_3^\pm(b-d,s,\lambda)=e^{\pm i\lambda p(b-d)}\chi_3^\pm(b-d,s,\lambda)\nn\\
0\leq d\leq b,\;\;\;\;0<s<a
\label{A21}
\ee

For the SWF $\Psi^{as}(x,y,\lambda)$ we have:
\be
\Psi^{as}(x,y,\lambda)=\Psi_1^+(d,s,\lambda)+\Psi_3^+(b-d,s,\lambda)=
\Psi_1^-(b-d,s,\lambda)+\Psi_3^-(d,s,\lambda)\nn\\
(x,y)=(s,d)
\label{A22}
\ee
together with the following identifications:
\be
\Psi_3^\pm(b-d,s,\lambda)=\Psi_1^\mp(d,s,\lambda)\nn\\
\Psi_1^+(0,s,\lambda)=-\Psi_1^-(0,s,\lambda)\nn\\
\Psi_3^+(0,s,\lambda)=-\Psi_3^-(0,s,\lambda)
\label{A23}
\ee
so that:
\be
\chi_3^\pm(b-d,s,\lambda)=e^{\mp i\lambda pb}\chi_1^\mp(d,s,\lambda)\nn\\
\chi_1^+(0,s,\lambda)=-\chi_1^-(0,s,\lambda)\nn\\
\chi_3^+(0,s,\lambda)=-\chi_3^-(0,s,\lambda)
\label{A24}
\ee

It follows from the last equations that:
\be
\chi_3^+(0,s,\lambda)=-\chi_3^-(0,s,\lambda)=-e^{i\lambda pb}\chi_1^+(b,s,\lambda)
\label{A25}
\ee
so that:
\be
\chi_3^+(b,s,\lambda)=-e^{i\lambda pb}\chi_1^+(2b,s,\lambda)
\label{A25a}
\ee
and further:
\be
\chi_1^+(0,s,\lambda)=-\chi_1^-(0,s,\lambda)=-e^{i\lambda pb}\chi_3^+(b,s,\lambda)=
e^{2i\lambda pb}\chi_1^+(2b,s,\lambda)=e^{2i\lambda pb}\chi_1^+(0,s,\lambda)
\label{A25b}
\ee
since $\chi_1^+(2b,s,\lambda)=\chi_1^{+,cont}(0,s,\lambda)=\chi_1^+(0,s,\lambda)$.

Therefore:
\be
e^{2i\lambda pb}=1
\label{A26}
\ee
and
\be
\Psi^{as}(x,y,\lambda)=e^{i\lambda pd}\chi_1^+(d,s,\lambda)+e^{-i\lambda pd}e^{i\lambda pb}\chi_3^+(b-d,s,\lambda)=\nn\\
e^{i\lambda pd}\chi_1^+(d,s,\lambda)- e^{-i\lambda pd}\chi_1^+(2b-d,s,\lambda)
\label{A27}
\ee

We have of course $\Psi^{as}(x,0,\lambda)=\Psi^{as}(x,b,\lambda)=0$ by construction. But we have to have also
$\Psi^{as}(0,y,\lambda)=\Psi^{as}(a,y,\lambda)=0$, i.e. we have to have:
\be
e^{i\lambda pd}\chi_1^+(d,0,\lambda)- e^{-i\lambda pd}\chi_1^+(2b-d,0,\lambda)=0\nn\\
e^{i\lambda pd}\chi_1^+(d,a,\lambda)- e^{-i\lambda pd}\chi_1^+(2b-d,a,\lambda)=0
\label{A28}
\ee

Therefore from \mref{A28} in the limit $\lambda\to\infty$ we have:
\be
e^{i\lambda pd}\chi_{1,0}^+(0)- e^{-i\lambda pd}\chi_{1,0}^+(0)=0\nn\\
e^{i\lambda pd}\chi_{1,0}^+(a)- e^{-i\lambda pd}\chi_{1,0}^+(a)=0
\label{A28a}
\ee
so that
\be
\chi_{1,0}^+(0)=\chi_{1,0}^+(a)=0
\label{A29}
\ee

Next let us invoke the second of the equations \mref{12z} to get in the considered case for $d=2b$:
\be
\chi_{1,1}^+(2b,s)=\chi_{1,1}^+(0,s)+i\frac{b}{p}\ll(\frac{d^2\chi_{1,0}^+(s)}{ds^2}+2E_1\chi_{1,0}^+(s)\r)
\label{A30}
\ee
so that
\be
\frac{d^2\chi_{1,0}^+(s)}{ds^2}+2E_1\chi_{1,0}^+(s)=0
\label{A31}
\ee
since $\chi_{1,1}^+(2b,s)=\chi_{1,1}^+(0,s)$.

The obvious solution of the last equation satisfying the boundary conditions \mref{A29} is:
\be
\chi_{1,0}^+(s)=A_0\sin(\sqrt{2E_1}s)\nn\\
\sqrt{2E_1}a=m\pi,\;\;\;\;\;m=1,2,...
\label{A32}
\ee

Coming back to the second of the equations \mref{12z} we can conclude that $\chi_{1,1}^+(d,s)$ again is independent of
$d$. Passing next to the third of the equations \mref{12z} and repeating arguments similar to those which led us to
\mref{A30}-\mref{A31} we get the following explicit dependence of $\chi_{1,1}^+(s)$ on $s$:
\be
\chi_{1,1}^+(s)=A_1\sin(\sqrt{2E_1}s)+B_1\cos(\sqrt{2E_1}s)+\frac{E_2A_0s}{\sqrt{2E_1}}\cos(\sqrt{2E_1}s)
\label{A33}
\ee

The boundary conditions $\chi_{1,1}^+(0)=\chi_{1,1}^+(a)=0$ enforce however $B_1=E_2=0$.

Using again \mref{12z} and the inductive arguments we come to the conclusion that $\chi_{1}^+(d,s,\lambda)$ is
$d$-independent and coefficients of its semiclassical series have the form:
\be
\chi_{1,k}^+(s)=A_k\sin(\sqrt{2E_1}s),\;\;\;\;\;\;k=0,1,...
\label{A34}
\ee
so is the form of $\chi_{1}^+(s,\lambda)$ itself, i.e.
\be
\chi_{1}^+(s,\lambda)=A(\lambda)\sin(\sqrt{2E_1}s)
\label{A35}
\ee

Therefore coming back to \mref{A27} we get:
\be
\Psi^{as}(x,y,\lambda)=(e^{i\lambda pd}- e^{-i\lambda pd})\chi_1^+(s,\lambda)=
2iA(\lambda)\sin(\lambda pd)\sin(\sqrt{2E_1}s)
\label{A36}
\ee
which again is the result of the previous way of the rectangular billiard energy quantization.

But now the energy $E$ is given by the (finite) semiclassical series:
\be
E=\fr p^2+\frac{E_1}{\lambda^2}=\fr\ll(\ll(\frac{n\pi}{\lambda b}\r)^2+\ll(\frac{m\pi}{\lambda a}\r)^2\r),\;\;\;\;\;m,n=1,2,...
\label{A37}
\ee

Therefore we get a surprising result that SWF's in the rectangular billiards can be built equivalently by the following
two ways:
\begin{enumerate}
\item on the skeletons which rays are inclined to the billiard sides with corresponding angles defined by the quantization
conditions \mref{A14} - in this case the $\chi$-coefficients of the SWF's are simply constant on the rectangular boundary;
\item on one of the two skeletons which rays are perpendicular to one of the rectangular sides - in this case the
corresponding $\chi$-coefficients vary along the sides perpendicular to the rays and vanish on the sides parallel to
them.
\end{enumerate}

But a more important conclusion which follows from the results got in this section is that a relation between the form
of skeletons and the SWF's and the corresponding energy spectrum which seemed to be suggested by the first part of the
section is rather illusory since a full description of these
quantities can be also obtained using a single skeleton only at least in the case of the rectangular billiards.

\subsection{Broken rectangular billiards}

\hskip+2em By a broken rectangular billiards we understand the one on Fig.7, i.e. with some number of rectangular
bays and peninsulas.
\begin{figure}
\begin{center}
\psfig{figure=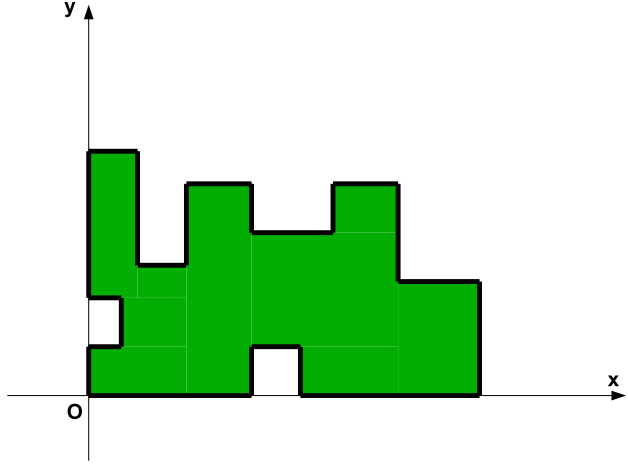,width=12cm}
\caption{An "arbitrary" broken rectangular billiards}
\end{center}
\end{figure}

To illustrate the way of energy quantization in such billiards we shall consider first the one
with a single peninsula shown in Fig.8. The corresponding procedures are analogous to the previous ones. One needs to
construct additional four bundles relating with the peninsula sides.
\begin{figure}
\begin{center}
\psfig{figure=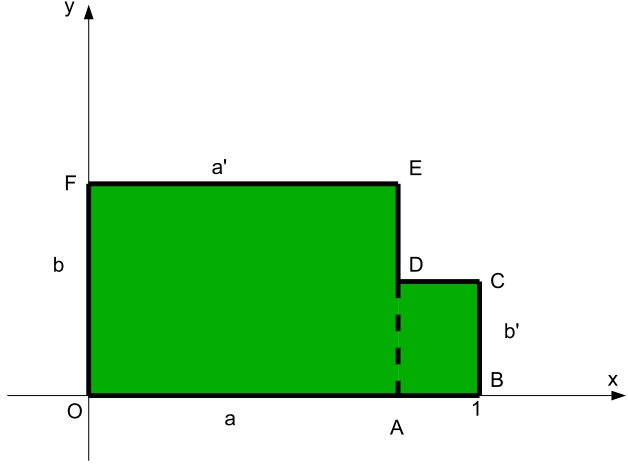,width=12cm}
\caption{A single bay rectangular billiards}
\end{center}
\end{figure}

Although both the previous methods of the sections
5.1 and 5.2 are equivalent leading us to the same results the method of sec.5.2 seems however to be simpler and more
instructive in applications to more complicated cases of billiards.

We assume from the very beginning that all the sides of the billiards from Fig.10, i.e. $a=1,b,a',b'$
are commensurate. This assumption
can always be satisfied even if the sides are expressed by irrational number just by respective approximations of the
latter by rational ones with arbitrary accuracies. We proceed as follows.

We construct two vertical skeletons. One for the rectangle $OAEF$ and the second for $ABCD$.

For the first skeleton according to \mref{A36} we get the following SWF:
\be
\Psi_1^{as}(x,y,\lambda)=A_1\sin(\lambda py)\sin(\sqrt{2E_1}x)\nn\\
0\leq x<a',\;0\leq y\leq b\nn\\
\lambda pb=m\pi,\;\;\;\;\;m=1,2,...\nn\\
\sqrt{2E_1}a'=n\pi,\;\;\;\;\;n=1,2,...
\label{A36a}
\ee
while for the second:
\be
\Psi_2^{as}(x,y,\lambda)=A_2\sin(\lambda p'y')\sin\ll(\sqrt{2E_1'}(1-x')\r)\nn\\
a'<x'\leq 1,\;0\leq y'\leq b'\nn\\
\lambda p'b'=k\pi,\;\;\;\;\;k=1,2,...
\label{A36b}
\ee

To get $\Psi_2^{as}(x,y,\lambda)$ for the total rectangular we have to match both the previous ones on the
segment $AD$ of Fig.8. Since this matching has to be valid for $0\leq y'\leq b'$ then as it follows easily we have
to have $p=p'$ so that $E_1=E_1'$ because of \mref{A37}, i.e. the vertical wave lengths of both the matched solutions
have to be the same. Therefore we get:
\be
A_1\sin(\sqrt{2E_1}a')=A_2\sin\ll(\sqrt{2E_1}(1-a')\r)\nn\\
A_1\cos(\sqrt{2E_1}a')=-A_2\cos\ll(\sqrt{2E_1}(1-a')\r)
\label{A36c}
\ee
so that $\sqrt{2E_1}=l\pi,\;l=1,2,...,$ and $A_2=(-1)^{l+1}A_1$.

Therefore the procedure leads us to the following quantization conditions for the energy $E_{nm}$:
\be
\frac{1}{n_0}=n\frac{\Lambda_{x}}{2},\;\;\;\;\;\;\;\;\frac{b}{m_0}=m\frac{\Lambda_{y}}{2}\nn\\
\Lambda_{x}=\frac{2\pi}{\sqrt{2E_1}},\;\;\;\;\;\;\;\;\Lambda_{y}=\frac{2\pi}{\lambda p}\nn\\
E_{nm}=\fr p^2+\frac{2E}{\lambda^2}=\frac{2\pi^2}{\lambda^2}\ll(\frac{1}{\Lambda_{x}^2}+\frac{1}{\Lambda_{y}^2}\r)=
\frac{\pi^2}{2\lambda^2}\ll(n^2n_0^2+\frac{m^2m_0^2}{b^2}\r)\nn\\
m,n=1,2,...
\label{A38}
\ee
where $\Lambda_{x},\;\Lambda_{y}$ are the wave lengths of rays in the
horizontal and vertical skeletons respectively and $n_0,m_0$ are the smallest integers satisfying $l_0=n_0a'$
and $k_0b=m_0b'$ where $l_0,k_0$ are also integers.

The respective SWF's are the following:
\be
\Psi_{nm}^{as}(x,y,\lambda)=\ll\{\ba{lr}
A\sin\frac{2\pi x}{\Lambda_{x}}\sin\frac{2\pi y}{\Lambda_{y}}=
A\sin(nn_0\pi x)\sin\ll(\frac{mm_0}{b}y\r)&\;\;\;\;\;\;(x,y)\in D_{br}\\
0&\;\;\;\;\;\;(x,y)\notin D_{br}
\ea\r.
\label{A38a}
\ee
where $D_{br}$ denotes the domain of the $x,y$-plane occupied by the broken rectangular of Fig.8.

One can easily realize that the last results can be easily generalized to any broken rectangular
billiards. A little bit surprising is that the formulae \mref{A38} for the energy and \mref{A38a} for the wave functions
remain unchanged for any such a billiards while a number of conditions the wave lengths $\Lambda_{x}$ and $\Lambda_{y}$
have to satisfy filling the vertical and horizontal skeletons by integer numbers of their halves is increasing
respectively to numbers of bays and peninsulas forming the sides od such billiards.

\section{The simplest SWF's solutions for polygons and Bunimovich billiards}

\hspace{15pt}The semiclassical quantization of the rectangular billiards in sec.5.2 shows that it is possible to
quantize semiclassically in the similar way some specific configurations of the SWF's in more complicated billiards
also classically chaotic such as polygons or Bunimovich billiards. Such easy opportunities appear if billiards
to be considered possess boundaries which allow us for easy constructions of corresponding skeletons. Some simple
examples of such billiards are provided by a parallelogram, a trapezium, a pentagon shown in Fig.9
\begin{figure}
\begin{center}
\psfig{figure=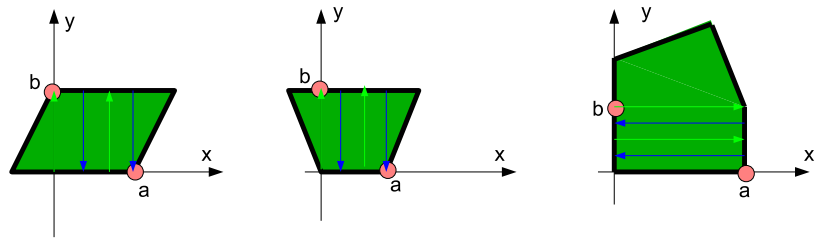,width=12cm}
\caption{A parallelogram, a trapezium and a pentagon billiards with the bouncing mode skeletons}
\end{center}
\end{figure}
or by the Bunimovich stadium - Fig.10,
\begin{figure}
\begin{center}
\psfig{figure=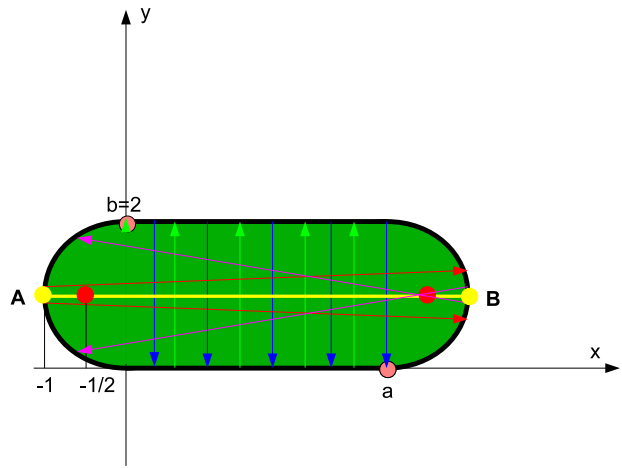,width=12cm}
\caption{The Bunimovich billiards with the bouncing mode skeleton and with the scar generating bundles associated with
the horizontal periodic orbit}
\end{center}
\end{figure}
and its simple generalizations - Fig.11.
\begin{figure}
\begin{center}
\psfig{figure=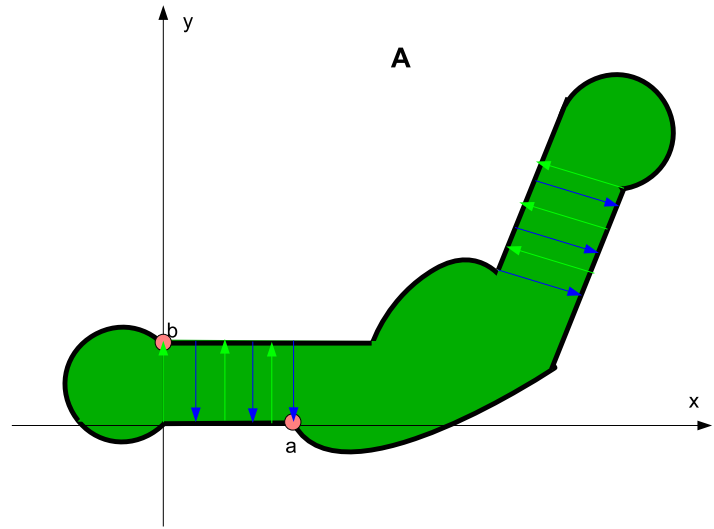,width=12cm}\\
\psfig{figure=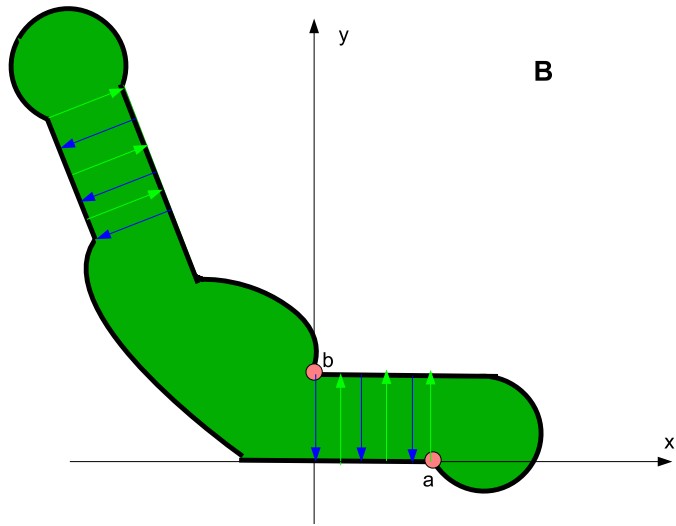,width=12cm}
\caption{A simple generalization of the Bunimovich billiards with the corresponding bouncing mode skeletons}
\end{center}
\end{figure}

The corresponding bundle skeletons are shown on the respective figures.

The constructions of the SWF's corresponding to the skeletons shown on the figures coincide with those for
the rectangular billiards and therefore these SWF's are the following:
\be
\Psi^{as,+}(x,y,\lambda)=A\sin(\lambda p_1x)\sin(\lambda p_2y),\nn\\
E=\fr\ll(p_1^2+p_2^2\r)\nn\\
\lambda p_1a=m\pi,\;\;\;\lambda p_2b=n\pi,\;\;\;\;m,n=1,2,...
\label{A40}
\ee
for the skeletons perpendicular to the $x$-axis and defined in the respective rectangles $a\times b$ of the figures
while outside the skeletons $\Psi^{as,+}(x,y,\lambda)=0$, i.e. those regions are classically "forbidden" and the
corresponding wave functions vanish there exponentially if $\lambda\to\infty$.
.

For the Bunimovich-like billiards of Fig.11 it should be clear that bouncing modes can be excited independently in
each of its rectangular part but also simultanuously in both parts tunning the modes respectively.

Formulae like \mref{A40} have been suggested by Burq and Zworski
\cite{15} by completely different but mathematically rigorous arguments.

\section{Scars and periodic orbits}

\hspace{15pt}It was shown in the previous section that the idea of the skeletons seems to be effective in solving some simple situations of
quantum phenomena related semiclassically with the chaotic dynamics. Nevertheless obvious difficulties in effective constructions of skeletons
in the cases of chaotic dynamics seems to limit seriously its applications. Despite this one can try to understand with it the scar phenomena
noticed by Heller \cite{10} and investigated by the latter author and others \cite{25,20,15,16}. Although in examples discussed below we do not construct closed skeletons but
yet supposing their existence in these cases allows us at least half-qualitatively to understand the scar phenomena and to make some predictions
for their existence in some cases of billiards and their absence in others.

Consider for example the Bunimovich stadium billiards of Fig.10 and the isolated unstable horizontal periodic orbit linking
the top points A and B
of the stadium. It is known \cite{10} that the there is a mode for which its wave function takes significantly larger values around the orbit than
far away of it, i.e. the orbit signals its existence by such a "scar". This and similar scar phenomena have not got their full description although there
were many efforts and approaches to do it \cite{25,20,15,16}.

Of course it is not easy not only to construct a skeleton carrying such a mode but even to prove its existence. Nevertheless assuming the latter we
can expect that such a skeleton will be symmetric horizontally and vertically as well as its bundles will contain the horizontal periodic orbit.

Consider one of such bundles emerging from a vicinity of the top point A on Fig.10. Even if its rays are divergent with respect to the periodic orbit the rays which are very close to it are also almost
parallel to it, i.e. after the reflection by the opposite semicircle of the billiards they are transformed into convergent bundle focused close to
the focal point of the reflecting semicircle. It means that  central parts of almost all such bundles of the skeleton containing the horizontal periodic
orbit have to be convergent and have to pass close to focal points of the semicircles.

Assume the radii of the semicircles to be equal to $1$ while according to Fig.10  the flat parts of the stadium have the
length $a$ each. It is clear that each such a convergent bundle will
generate after the reflection a new bundle the central part of which is again convergent. On this new bundle a new $\chi$-factor defined on it will
however be weakened according to \mref{22b}
by the factor $\ll|\frac{\p h(s;u,l)}{\p s}\r|^{-\fr}\approx (2a+3)^{-\fr}$. Therefore if the starting bundle has the factor $\chi_0$ as its
zeroth order semiclassical approximation then after
$n$ subsequent reflections it will be weakened by the factor $(2a+3)^{-\fr n}$. However the central parts of all the reflected bundles will
remain close to the periodic orbit so that SWF's defined on them will interfere in an infinite number of them close to the orbit.

Easy calculations of the contributions coming from all such bundles lead us to the following form of the regular scarring
part of the semiclassical wave function on the horizontal periodic orbit of Fig.10 in the JWKB approximation:
\be
\Psi_{scar}^{JWKB}(x,1)=\nn\\
\frac{e^{i\lambda p(a+2)}}{1-q^2e^{2i\lambda p(a+2)}}\ll(\ll(2x+1\r)^{-\fr}e^{-i\lambda p(a-x+1)}-
\ll(2(a-x)+1\r)^{-\fr}qe^{+i\lambda p(a-x+1)}\r)\chi_0(1)\nn\\
-1\leq x\leq a+1,\;x\neq-\fr,a+\fr,\;\;\;\;q=(2a+3)^{-\fr}
\label{A41}
\ee
where $\arg(2x+1)=\pi$ for $x<-\fr$ and the point $x=-\fr$ is avoided clockwise and above it when $x$ moves to the values
larger than $-\fr$ and the point $A$ has been chosen as the initial one for the $s$-variable so that $s\approx y-1$ in
a vicinity of this point.

Certainly the above contribution is not a unique one ($\Psi_{scar}^{JWKB}(x,1)$ does not vanish for $x=-1,a+1$, see our
comment below). There
are still infinitely many bundles contributing to the SWF coming from the parts of the
bundles discussed above scattered away of their central parts. However it seems reasonable to assume that just such bundles scattered upon the
\begin{figure}
\begin{center}
\psfig{figure=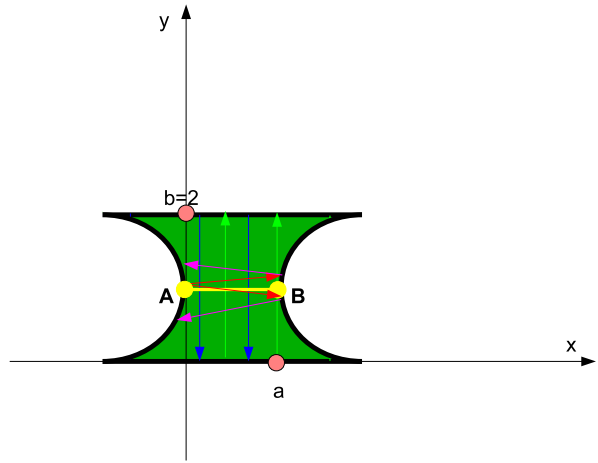,width=12cm}
\caption{The anti-Bunimovich billiards with the bouncing mode skeleton deprived of the scar generating bundles associated with
the horizontal periodic orbit}
\end{center}
\end{figure}
whole billiards interfere everywhere
chaotically, i.e. typically as in the cases without scars and satisfying Berry's conjecture \cite{19}. Some order in such interfering can be organized as we can try to argue by the presence
of periodic orbits supported by the focusing properties of the billiards boundary in vicinities of top points, i.e.
the points where periodic orbits touch the billiards boundary.

The last condition, i.e. the focusing properties of the billiards boundary at the top points seems to be essential for
the scar phenomenon to appear. One can be convinced of its necessity
by considering an "anti-Bunimovich" stadium, i.e. the stadium which semicircular parts instead of being concave are
convex for the billiards, see Fig.12. The horizontal periodic orbit still exists but a skeleton with the properties discussed earlier certainly does not, i.e. the
bundles containing this periodic orbit can be only scattered so that the only rays of these bundles which come back close to the periodic orbit is
the orbit itself. All the remaining rays even close to the periodic orbits are scattered away passing by the virtual
focus points of the convex semicircles.

The results of Barnett's calculations demonstrated by Sarnak \cite{21,27} seem
also to confirm this conclusion.

Let us note however that the above analysis is not sufficient to estimate the energy corresponding to the mode described above
even its lowest JWKB approximation investigating only a vicinity of the horizontal periodic orbit . This is because
after every reflection the corresponding half of the horizontal periodic orbit belongs to a new bundle, i.e. this
orbit is in fact not closed on the skeleton considered and moreover it does not end in the initial bundle (this is why
it is not periodic inside the skeleton). Because of that this classically periodic orbit cannot be used to write on it
"the last quantization condition" of the form \mref{23c}. In fact such a single periodic orbit possibility of
determining the energy
would question the role of the eigenfunction boundary conditions in forming its corresponding eigenvalue.
\begin{figure}
\begin{center}
\psfig{figure=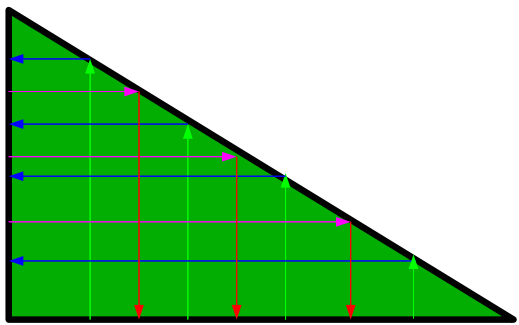,width=12cm}
\caption{The example of the four bundles skeleton in the right angle triangular billiards on which none semiclassical
eigenfunction can be defined}
\end{center}
\end{figure}

\section{Summary and conclusions}

\hspace{15pt}We have shown in this paper that it is possible to formulate the semiclassical description of the quantum billiards eigenvalue problems
in the semiclassical wave function language. We have used in principle the approach of Maslov and Fedoriuk \cite{4}
modifying it
however by the way of moving through caustics. The corresponding procedure has been described in App.B. where the idea
of continuing semiclassical wave functions by coustic singularities on the complex $t$-plane was developed. It was shown
there a close relation between a signature of a SWF and a path along which the SWF has to avoid a caustic
singularity on the $t$-plane.

We have also
modified the Maslov and Fedoriuk procedure by constructing respective Lagrange manifolds not as smooth,
deprived of boundaries tori-like surfaces but rather in the form of skeletons. The skeleton idea has appeared to be sufficiently
flexible to gather quantum systems at least in principle uniformly
independently of the kind of their semiclassiclal limits, i.e. whether these limits are integrable or chaotic.
Nevertheless in the last cases
constructions of the corresponding skeletons seem to be rather difficult except these obvious chaotic billiards boundary configurations which
have been considered in this paper.

It is worth to stress also at this summary that the forms of skeletons as a set of ray bundles satisfying the geometrical optics law of
the mirror-like reflection from the billiards boundary was not a matter of a free choice but a necessity of satisfying
the boundary conditions by the SWF's as it was shown in App.A.

No less important was establishing that in the semiclassical
calculations the zeroth order terms of SWF's are classical integrals of motion. This has allowed us to close the
corresponding calculations.

Using known examples of the classically integrable billiards such as the circular and rectangular ones and
the broken rectangular billiards as some variants of the rectangular ones we have demonstrated the effectiveness of
the skeleton method in the quantization of the systems mentioned. In particular the semiclassical calculations of energy
levels performed in sec.5 have developed a new algorithm of finding these levels to any order in Planck constant for the
case of the circular billiards (i.e. the cylindrical infinite well).

Applications of the method to other classically integrable billiards systems also in higher than two dimensions are in
preparations.

We have shown also how the skeleton method allows us to describe almost trivially the bouncing ball modes in the
Bunimovich billiards and similar ones.

We have also discussed a possible explanation of the scar phenomena by the skeleton idea showing a close relation of
these phenomenta to focusing properties of billiards boundary in vicinity of the top points of the periodic orbits
carrying scars. A lack of focussing properties by the billiards boundary in vicinity of the top points should exclude
the existence of scars around the corresponding periodic orbits.

On the other hand it should be stressed that constructing a desired skeleton does not guarantee a construction on
it a SWF satisfying desired necessary boundary and other
conditions. A simple example of such a situation is a rectangular triangel billiards shown on Fig.13. There is a skeleton collected of
four bundles as shown in the figure. It is clear however that none SWF vanishing on the boundary of the billiards can
be built on this skeleton.

\section*{Appendix A}

\hspace{15pt} In this appendix we are going to show, that the geometrical optics rule of reflections of rays off the
billiards boundary is a consequence of demands of vanishing on the boundary of the linear combination \mref{19} accompanied by the conditions
\mref{19a} and \mref{20}. Namely consider the following superposition of SWF's:
\be
\Psi_k^{as}(x,y;u,l;\lambda)=\Psi_{k,1}^{\sigma_1}(d_1,s_1;u,l;\lambda)+\Psi_{k,2}^{\sigma_2} (d_2,s_2;u,l;\lambda)=\nn\\
J_{k,1}^{-\fr}(d_1,s_1;u,l)e^{i\sigma_1k\ll(d_1+\int_0^{s_1}\cos\alpha_{k,1}(s';u,l)ds'\r)}
\chi_{k,1}^{\sigma_1}(d_1,s_1;u,l;\lambda)+\nn\\
J_{k,2}^{-\fr}(d_2,s_2;u,l)e^{i\sigma_2k\ll(d_2+\int_0^{s_2}\cos\alpha_{k,2}(s';u,l)ds'\r)}
\chi_{k,2}^{\sigma_2}(d_2,s_2;u,l;\lambda)
\label{19b}
\ee
with
\be
{\bf r}\equiv[x,y]={\bf r}_{k,1}(d_1,s_1;u,l)={\bf r}_0(s_1)+{\bf d}_{1}(s_1;u,l)=\nn\\{\bf r}_{k,2}(d_2,s_2;u,l)=
{\bf r}_0(s_2)+{\bf d}_{2}(s_2;u,l)\nn\\
{\bf r}_{k,1}(d,s;u,l)\in B_k(u,l),\;\;\;\;\;{\bf r}_{k,2}(d,s;u,l)\in B_k'(u,l),\;\;\;\;B_k(u,l)\neq B_k'(u,l)
\label{19c}
\ee
i.e. the SWF's $\Psi_{k,1}^{\sigma_1} (d_1,s_1;u,l;\lambda)$ and $\Psi_{k,2}^{\sigma_2}(d_2,s_2;u,l;\lambda)$ are defined
respectively on the bundles $B_k(u,l)$ and $B_k'(u,l)$ with $D_k(u,l)\cap D_k'(u,l)\neq\oslash$ interfering in the
crossing point $[x,y]$ of two rays ${\bf r}_{k,1}(d_1,s_1;u,l)$ and ${\bf r}_{k,2}(d_2,s_2;u,l)$ belonging to the
respective bundles.

Therefore the condition for $\Psi_k^{as}(x,y,\lambda)$ to vanish on $A_k(u,l)$ is:
\be
J_{k,1}^{-\fr}(0,s)e^{ik\sigma_1\int_0^s\cos\alpha_{k,1}(s';u,l)ds'}\chi_{k,1}^{\sigma_1}(s;u,l;\lambda)+\nn\\
J_{k,2}^{-\fr}(0,s)e^{ik\sigma_2\int_0^s\cos\alpha_{k,2}(s';u,l)ds'}\chi_{k,2}^{\sigma_2}(s;u,l;\lambda)=0\nn\\
{\bf r}_0(s)\in A_k(u,l)
\label{21}
\ee

Because of the $k$-dependence the last relation can be satisfied if and only if:
\be
\sigma_1\int_0^s\cos\alpha_{k,1}(s';u,l)ds'=\sigma_2\int_0^s\cos\alpha_{k,2}(s';u,l)ds',
\;\;\;\;\;\;{\bf r}_0(s)\in A_k(u,l)
\label{21a}
\ee

It is easy to see however that there are only two solutions of the last condition:
\be
\alpha_{k,1}(s;u,l)\equiv\alpha_{k,2}(s;u,l)\;\;\;\;\;\;\;\; for\;\;\;\;\;\;\;\;\sigma_1=\sigma_2 \nn\\
\alpha_{k,1}(s;u,l)\equiv\pi-\alpha_{k,2}(s;u,l)\;\;\;\;\;\;\;\; for\;\;\;\;\;\;\;\; \sigma_1=-\sigma_2\nn\\
{\bf r}_0(s)\in A_k(u,l)
\label{21c}
\ee

The first solutions are however uninteresting identifying the bundles in a given segment and
consequently leading to the solutions vanishing identically on $A_k(u,l)$.

Putting $\alpha_{k,1}(s;u,l)\equiv\alpha(s;u,l)$ and $\sigma_1=-\sigma$ we get from the second solution and from \mref{21}:
\be
\chi_{k,2}^{-\sigma}(s;u,l;\lambda)=-\chi_{k,1}^{\sigma}(s;u,l;\lambda)\equiv\chi_k(s;u,l;\lambda)\nn\\
{\bf r}_0(s)\in A_k(u,l)
\label{21b}
\ee
so that the combination \mref{19} becomes:
\be
\Psi_k^{as}(x,y;u,l;\lambda)=\Psi_{k,1}^{\sigma}(d_1,s_1;u,l;\lambda)-\Psi_{k,2}^{-\sigma} (d_2,s_2;u,l;\lambda)=\nn\\
J_{k,1}^{-\fr}(d_1,s_1;u,l)e^{i\sigma k\ll(d_1+\int_0^{s_1}\cos\alpha(s';u,l)ds'\r)}
\chi_{k,1}^{\sigma}(d_1,s_1;u,l;\lambda)-\nn\\
J_{k,2}^{-\fr}(d_2,s_2;u,l)e^{-i\sigma k\ll(d_2+\int_0^{s_2}\cos\alpha(s';u,l)ds'\r)}
\chi_{k,2}^{-\sigma}(d_2,s_2;u,l;\lambda)\nn\\
{\bf r}_0(s)\in A_k(u,l)
\label{21d}
\ee
where $\chi_k{\sigma}(d,s;u,l;\lambda),\;\sigma=\pm,$ are given by \mref{13A} with
$\chi_k^{\sigma}(0,s;u,l;\lambda)\equiv\chi_k(s;u,l;\lambda)$.

The last result shows that $\Psi_k^{as}(x,y;u,l;\lambda)$ vanishing on $A_k(u,l)$ has to be
represented semiclassically by a
combination of at least two SWF's of opposite signatures and such that if $\Psi_{k,1}^{\sigma}(d,s;u,l;\lambda)$ is
defined on the bundle $B_k(u,l)$
then the second SWF $\Psi_{k,2}^{-\sigma}(d,s;u,l;\lambda)$ has to be defined on the bundle $B_k^A(u,l)$.

\section*{Appendix B}

\hspace{15pt}In this appendix we explore the well known properties of the circular billiards to establish the way of
avoiding the singular caustic points on the $t$-plane as well as to confirm the way given by \mref{22} by which two
SWF's have been matched.

The circular billiards is of course the well known case of the two dimensional infinitely deep cylindrical potential
well which energy spectrum is easily obtained by solving the stationary Schr\"odinger equation (SSE) for this case by
the variable separation method performed in the cylindrical coordinates. In the latter coordinates the radial part of the
SSE is the following:
\be
\Psi''(r)+\frac{1}{r}\Psi'(r)+\lambda^2\ll(E-\frac{m^2}{\lambda^2r^2}\r)\Psi(r)=0
\label{24}
\ee
where the separation constant $m=0,\pm 1,...$, is the angular momentum quantum number, $\lambda=\hbar^{-1}$ and we
have put $2M=1$ where $M$ is the billiard ball mass.

After the substitution $\Psi(r)=\frac{\psi(r)}{r^\fr}$ we get:
\be
\psi''(r)+\lambda^2\ll(E-\frac{m^2-\frac{1}{4}}{\lambda^2r^2}\r)\psi(r)=0
\label{25}
\ee

Assuming the unit radius of the billiards we get the solutions for \mref{24} in the following forms \cite{8}:
\be
{\tilde\psi}_\pm(r)=q^{-\frac{1}{4}}e^{\pm i\lambda\int_1^rq^\fr dr'}{\tilde\chi}_\pm(r),\;\;\;\;r_+<r\leq 1\nn\\
q=p_r^2=E-\frac{m^2}{\lambda^2r^2}=\frac{E}{r^2}(r-r_+)(r-r_-),\;\;\;\;r_\pm=\pm\frac{|m|}{\lambda\sqrt{E}},
\;\;\;m\neq 0
\label{26}
\ee
which correspond to the fundamental solutions ${\tilde\psi}_+(r)$ and ${\tilde\psi}_-(r)$
defined in the respective sectors $1$ i ${\bar 1}$ of the Stokes graph of Fig.13 with ${\tilde\chi}_\pm(r)$ normed at the
corresponding sector infinities by ${\tilde\chi}_+(\infty_1)={\tilde\chi}_-(\infty_{\bar 1})=1$. The $r$-plane on this figure
has a cut from $r_+$ to $r_-$ on which $q^\fr$ changes its sign, while $q^{-\frac{1}{4}}$ gets the factor $\pm i$
depending on the cut crossing directions ($+i$ for crossing it upwards, $-i$ - for downwards).
\begin{figure}
\begin{center}
\psfig{figure=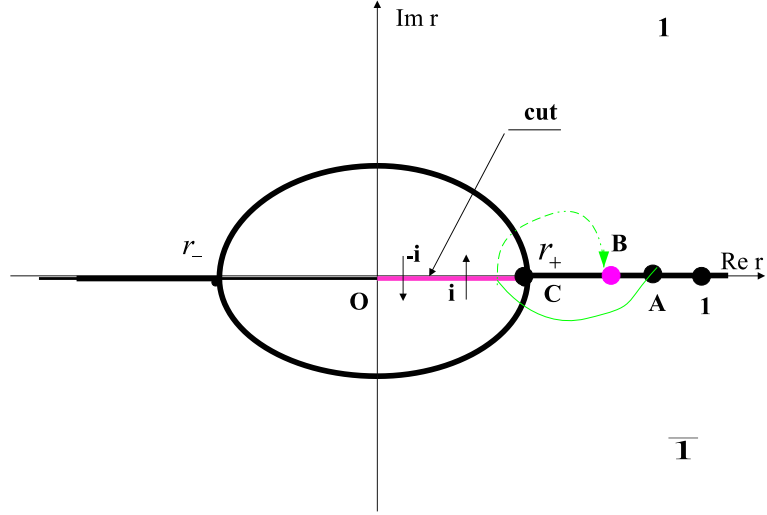,width=12cm}
\caption{The Stokes graph for the cylindrical billiards}
\end{center}
\end{figure}

The Stokes graphs formalism corresponding to \mref{26} easily provides us with the energy spectrum of the billiards. For this
we take the wave function ${\tilde\psi}(r)$ corresponding to this case as a linear combination
\be
{\tilde\psi}(r)={\tilde\psi}_+(r)+c{\tilde\psi}_-(r)
\label{26a}
\ee
given on the segment $AC$ where $c$ is defined by the condition ${\tilde\psi}(1)=0$.

It is easy to see that then:
\be
c=-\frac{{\tilde\chi}_+(1)}{{\tilde\chi}_-(1)}
\label{26b}
\ee

Next the combination \mref{26a} has to be continued to the segment $OC$
vanishing on it when $r\to 0$. This means however that up to a constant it should be identified with the fundamental
solution ${\tilde\psi}_0(r)=q^{-\frac{1}{4}}e^{-i\lambda\int_{r_+}^rq^\fr dr'}{\tilde\chi}_0(r)$, taken for example in this form above the cut $OC$
attached to the sector 0 on Fig.13, i.e. for $r+i\epsilon,\;0<r<r_+$.

According to the well known rules the latter can be expressed as the following linear combination
of ${\tilde\psi}_\pm(r)$ \cite{8}:
\be
{\tilde\psi}_0(r)=-i{\tilde\chi}_{0\to{\bar 1}}e^{i\lambda\int_{r_+}^1q^\fr dr'}{\tilde\psi}_+(r)+
{\tilde\chi}_{0\to 1}e^{-i\lambda\int_{r_+}^1q^\fr dr'}{\tilde\psi}_-(r)\nn\\
0<r<r_+
\label{27}
\ee
where ${\tilde\chi}_{0\to j}=\lim_{r\to\infty_j}{\tilde\chi}_0(r)$ with $\infty_j$ being the infinity point in the sector $j,\;j=1,{\bar 1}$.

It then follows from \mref{27} that:
\be
{\tilde\psi}(r)=\frac{i}{\chi_{0\to{\bar 1}}}e^{-i\lambda\int_{r_+}^1q^\fr dr'}{\tilde\psi}_0(r)
\label{28}
\ee
and
\be
c=i\frac{{\tilde\chi}_{0\to 1}}{{\tilde\chi}_{0\to{\bar 1}}}e^{-2i\lambda\int_{r_+}^1q^\fr dr'}
\label{29}
\ee

The equations \mref{26b} and \mref{29} together are just the quantization conditions for the energy with the spectrum
of the latter being degenerate with respect to the sign of $m$.

To make further a correspondence with sec.5 we shall use rather the solutions $\psi_\pm(r)\equiv{\tilde\chi}_\mp(1)
{\tilde\psi}_\pm(r)$, $\chi_\pm(r)\equiv{\tilde\chi}_\mp(1){\tilde\chi}_\pm(r)$ and $\psi(r)\equiv{\tilde\chi}_-(1){\tilde\psi}(r)$
in the remaining considerations. Therefore instead of \mref{26a} we get simply:
\be
\psi(r)=\psi_+(r)-\psi_-(r)
\label{26ab}
\ee
so that
and instead of \mref{29} we have:
\be
1=-i\frac{\chi_{1\to 0}}{\chi_{{\bar 1}\to 0}}e^{-2i\lambda\int_{r_+}^1q^\fr dr'}
\label{29a}
\ee
where $\chi_{1\to 0}={\tilde\chi}_-(1){\tilde\chi}_{1\to 0}$ and $\chi_{{\bar 1}\to 0}={\tilde\chi}_+(1){\tilde\chi}_{{\bar 1}\to 0}$
and where we have taken into account that $\chi_{0\to j}=\chi_{j\to 0},\;j=1,{\bar 1}$.

The results \mref{26b}-\mref{29a} obtained above are exact.

Consider now the semiclassical limits of \mref{26b}-\mref{29a}, i.e. when $\lambda\to+\infty$. It should be
stressed that the fundamental solution
formalism allows us to do such a passage and in every of the above formula this limit is well defined. In particular all exponentially small
contributions have to be neglected. In this way if $r$ is in sector $0$ such an exponentially small is the solution
${\tilde\psi}_0(r)$ so that taking the semiclassical limit in \mref{27} we get:
\be
0=-i\chi_{{\bar 1}\to 0}^{as}e^{+i\lambda\int_{r_+}^1q_{as}^\fr dr'}\psi_+^{as}(r)+
\chi_{1\to 0}^{as}e^{-i\lambda\int_{r_+}^1q_{as}^\fr dr'}\psi_-^{as}(r),\;\;\;\;\;\;\;0<r\leq r_+\nn\\
\psi_\pm^{as}(r)=q_{as}^{-\frac{1}{4}}e^{\pm i\lambda\int_1^rq_{as}^\fr dr'}\chi_\pm^{as}(r)
\label{30}
\ee
where $"as"$ denotes the semiclassical forms of the relevant quantities and in particular
$\chi_\pm^{as}(r),\;\chi_{{\bar 1}\to 0}^{as}(=\chi_{0\to 1}^{as})$ and the asymptotic form of
$\chi_{1\to 0}^{as}(=\chi_{0\to{\bar 1}}^{as})$ can be found in any of the references \cite{8} while
$q_{as}=E-\frac{m^2}{\lambda^2r^2}$ with $E=E_0+\frac{1}{\lambda^2}E_1+...$.

In fact the linear relation \mref{30} is equivalent to the following {\bf identity} valid in the semiclassical limit:
\be
\chi_+^{as}(r)\chi_{{\bar 1}\to 0}^{as}=\chi_{1\to 0}^{as}\chi_-^{as}(r),\;\;\;\;\;\;\;0<r\leq r_+
\label{33}
\ee
if the point $r,\;r\in OC$, is approached continuing $\chi_+^{as}(r)$ anticlockwise and $\chi_-^{as}(r)$ - clockwise around the point $C$ of Fig.13.

Taking also the semiclassical limit of the exact quantization condition \mref{29a} we get:
\be
1=-i\frac{\chi_{1\to 0}^{as}}{\chi_{{\bar 1}\to 0}^{as}}e^{-2i\lambda\int_{r_+}^1q^\fr dr'}
\label{29b}
\ee

From both \mref{30} and \mref{29b} it follows that:
\be
\psi_+^{as}(r)=\psi_-^{as}(r),\;\;\;\;\;\;\;0<r\leq r_+
\label{31}
\ee

Conversely if \mref{31} is valid then from the identity \mref{30} we get the semiclassical form \mref{29b} of the quantization
condition.

Therefore both the conditions \mref{29a} and \mref{31} are {\bf equivalent} to each other as
the semiclassical quantization conditions in the case considered.

We conclude therefore that the equation \mref{31} substitutes the quantization condition \mref{29b} in the semiclassical
limit, i.e. in the
quantized semiclassical limit both $\psi_+^{as}(r)$ and $\psi_-^{as}(r)$ has to coincide in the sector 0. But this
coincidence can be analytically continued to any (non singular) point of the $r$-plane. In particular it can be continued
back to the classically allowed segment $AC$. These can be done in two ways, by avoiding the singular point $C$ clockwise or
anticlockwise.

Before making the respective continuations let us note that the relation \mref{27} cannot be continued to the segment $AC$ of the Stokes graph of
Fig.13 not loosing its
connection with its asymptotic form \mref{30} while this asymptotic form itself can loosing however its connection to the original equation
\mref{27}. This new form of an equation substituting the original one \mref{27} can be obtained by the Borel resummation procedure.

Let us start therefore with the relation \mref{30} and \mref{31} to continue them to $AC$ around the point $C$ clockwise. We get:
\be
0=-i\chi_{{\bar 1}\to 0}^{as}e^{+i\lambda\int_{r_+}^1q_{as}^\fr dr'}\psi_+^{as}(r)+
\chi_{1\to 0}^{as}e^{-i\lambda\int_{r_+}^1q_{as}^\fr dr'}\psi_-^{as,cont}(r),\;\;\;\;\;\;\;0<r\leq r_+\nn\\
\label{30a}
\ee
and
\be
\psi_+^{as}(r)=\psi_-^{as,cont}(r),\;\;\;\;\;\;r_+<r\leq 1
\label{33aa}
\ee
where $\psi_-^{as,cont}(r)$ denotes $\psi_-^{as}(r)$ continued analytically clockwise around the
point $C$ from the segment $AC$ to the same segment.

If we choose the opposite path of continuation, i.e. anticlockwise one, then continuing
the relation \mref{31} we get:
\be
0=-i\chi_{{\bar 1}\to 0}^{as}e^{+i\lambda\int_{r_+}^1q_{as}^\fr dr'}\psi_+^{as,cont}(r)+
\chi_{1\to 0}^{as}e^{-i\lambda\int_{r_+}^1q_{as}^\fr dr'}\psi_-^{as,cont}(r),\;\;\;\;\;\;\;0<r\leq r_+\nn\\
\label{30b}
\ee
and
\be
\psi_-^{as}(r)=\psi_+^{as,cont}(r),\;\;\;\;\;\;r_+<r\leq 1
\label{33ba}
\ee
where $\psi_+^{as,cont}(r)$ denotes $\psi_+^{as}(r)$ continued analytically anticlockwise around the
point $C$ from the segment $AC$ to the same segment.

However, contrary to $OC$ the segment $AC$ lies now totally in the common part of the domains of Borel summability of $\psi_\pm(r)$ so that we have:
\be
\ll(\psi_\pm^{as}(r)\r)^{BS}=\psi_\pm(r),\;\;\;\;\;\;r_+<r\leq 1
\label{33ac}
\ee
where the superscript $BS$ means the Borel resummation of the relevant quantity.

But according to \mref{33aa} i \mref{33ba} we have instead:
\be
\ll(\psi_\pm^{as,cont}(r)\r)^{BS}=
\psi_\mp(r)\nn\\
r_+<r\leq 1
\label{33ad}
\ee
while by the Borel summing \mref{30a} and \mref{30b} we obtain respectively:
\be
\ll(\psi_-^{as,cont}(r)\r)^{BS}=i\frac{\chi_{{\bar 1}\to 0}}{\chi_{1\to 0}}e^{+2i\lambda\int_{r_+}^1q^\fr dr'}\psi_+(r)\nn\\
\ll(\psi_+^{as,cont}(r)\r)^{BS}=-i\frac{\chi_{1\to 0}}{\chi_{{\bar 1}\to 0}}e^{-2i\lambda\int_{r_+}^1q^\fr dr'}\psi_-(r)\nn\\
0<r\leq r_+
\label{30c}
\ee

If we now apply the Borel summed quantization conditions \mref{33ad} to the respective equations in \mref{30c} we immediately recover the exact
quantization condition \mref{29a}.

Let us summarize the above results.
\begin{itemize}
\item In the semiclassical limit the quantization condition for the circular billiards is the appropriate coincidence of the asymptotic expansions
of two fundamental solutions in the classically unallowed region (the segment $OC$). This coincidence is given by
\mref{31}.
\item In the classically allowed region the semiclassical quantization condition can be obtained by continuing
appropriately (clockwise or anticlockwise) one of the semiclassical solutions $\psi_\pm^{as}(r)$ from the segment $AC$
and back to it around the turning point $C$ and next by identifying the continued solution with the second one. The respective identifications are given by
\mref{33aa} and \mref{33ba}.
\item The semiclassical quantization conditions formulated in the classically allowed as well as
in unallowed regions can be Borel summable leading us to the exact ones;
\item The fundamental semiclassical solutions continued to the classically allowed region (the segment $AC$) by the
unallowed one (the segment $OC$) still represent semiclassical expansions of some fundamental solutions which by Borel
resummation can be identified appropriately (see the formulae \mref{30c}).
\item The semiclassical quantization condition formulated by \mref{33aa} and \mref{33ba} in the classically allowed region
(the segment $AC$) can be Borel summable providing us with the proper exact quantization conditions.
\end{itemize}

\subsection*{B.1  The fundamental solutions semiclassical analysis rewritten on the classical trajectories}

\hspace{15pt}Let us now rewrite the above results in the formalism of the SWF's defined on the classical trajectories
developed in sec.5.

To this goal let us assume the semiclassical expansion \mref{3a} for the energy $E$ so that the classical momentum
$p=\sqrt{2E_0}$, and associate with the solution \mref{26a} two bundle skeletons ${\bf B}$ and ${\bf B}_A$.

Both the skeletons consist of single ray bundles only. The skeleton ${\bf B}$ consists of the ray bundle $B(0,2\pi)$
the rays of which run with the
definite positive angular momentum $m,\;m=1,2,...,$ and with momenta ${\bf p}$ shown in Fig.2 while the angular momentum of
the skeleton ${\bf B}^A$ is equal to $-m$ and rays of its unique bundle $B^A(0,2\pi)$ have momenta ${\bf p}^A$ on Fig.2.
For both the momenta we have $|{\bf p}|=|{\bf p}^A|=p$.

Any point of the ring $(r,\Phi),\;r_+<r\leq 1,\;0\leq\Phi< 2\pi$, is crossed by two rays coming from different bundles
starting from the points $s_1$ and $s_2$ of the billiard boundary and having the respective angular momenta
equal to $p\cos\alpha=m/\lambda$ and $p\cos(\pi-\alpha)=-m/\lambda$ with $\cos\alpha=r_+$ (see Fig.2) while their
corresponding radial momenta are the same and equal to $p_r={\dot r}=\frac{p^2(t-t_c)}{r(t)}=
-\sqrt{E_0-\frac{p^2\cos^2\alpha}{r^2}}=-\sqrt{E_0-\frac{m^2}{\lambda^2r^2}}\equiv-q_0^\fr$ being negative
($t<t_c,\;t_c=\sin\alpha/p$).

Starting at the
moment $t=0$ from the boundary these rays meet each other at the point $(r,\Phi)$ in the moment $t$.

Each of the bundles defines a
transformation of coordinate $(r,\Phi)\to(d,s_i),\;i=1,2,$ by $r=r(d)=\sqrt{\cos^2\alpha+(\sin\alpha-d)^2},\;d=pt$ and
$\Phi=s_1+\alpha-\gamma=s_2-\alpha+\gamma<s_1+\alpha$ correspondingly to the bundle, see Fig.2.

The total original solution
\be
\Psi(r,\Phi)=\frac{\psi(r)}{r^\fr}e^{im\Phi}
\label{33bc}
\ee
corresponding to \mref{26ab} can be now rewritten by the new coordinates noticing that
$\int_1^rq_0^\fr dr'=-\int_1^rp_rdr'=
-\int_0^tp_r^2dt'=-\int_0^tp^2dt'+\int_0^t\frac{p_\Phi^2}{r^2}dt'=-p^2t+p^2\cos^2\alpha\int_0^t\frac{dt'}{r^2(t')}$ and
that
\be
\int_0^t\frac{dt'}{r^2(t')}=\ll.\frac{1}{\cos\alpha}\arctan\frac{p(t'-t_c)}{\cos\alpha}\r|_0^t=
\frac{1}{\cos\alpha}(\alpha-\gamma)
\label{31b}
\ee
for $t<t_c$ so that:
\be
\psi_\pm(r(t),\lambda)=q_0^{-\frac{1}{4}}e^{\mp i\lambda p^2t\pm i|m|(\alpha-\gamma)}\chi_\pm(t,\lambda)\nn\\
\chi_\pm(t,\lambda)=\ll(\frac{q}{q_0}\r)^{-\frac{1}{4}}
e^{\mp i\lambda\int_0^t\ll(q^\fr-q_0^\fr\r)q_0^\fr dt'}{\chi}_\pm(r(t),\lambda)=\nn\\
\ll(\frac{q}{q_0}\r)^{-\frac{1}{4}}
\exp\ll[\mp i\sum_{n\geq 0}E_{n+1}\lambda^{-n-1}\int_0^d\frac{1}{1+\ll(\frac{q}{q_0}\r)^\fr}dt'\r]
{\chi}_\pm(r(t),\lambda)\nn\\
\frac{q}{q_0}=1+\frac{\sum_{n\geq 1}E_n\lambda^{-n-1}}{q_0}
\label{31a}
\ee
and then we get:
\be
\Psi(r,\Phi)=\frac{\psi_+(r(t))-\psi_-(r(t))}{r^\fr(t)}e^{im\Phi}=\nn\\
(r^2q_0)^{-\frac{1}{4}}e^{-i\lambda p^2t+i|m|(\alpha-\gamma)+im\Phi}\chi_+(t,\lambda)-
(r^2q_0)^{-\frac{1}{4}}e^{+i\lambda p^2t-i|m|(\alpha-\gamma)+im\Phi}\chi_-(t,\lambda)
\label{32a}
\ee
where $(r^2q_0)^{-\frac{1}{4}}\equiv (r^2(t)q_0(t))^{-\frac{1}{4}}=(E_0(r^2(t)-r_+^2))^{-\frac{1}{4}}=
p^{-1}|t-t_c|^{-\fr}$.

From \mref{32a} we get further:
\be
\Psi(r,\Phi)=\ll\{\ba{lr}
(r^2q_0)^{-\frac{1}{4}}e^{-i\lambda p^2t+ims_2}\chi_+(t,\lambda)-
(r^2q_0)^{-\frac{1}{4}}e^{i\lambda p^2t+ims_1}\chi_-(t,\lambda),&m>0\\
(r^2q_0)^{-\frac{1}{4}}e^{-i\lambda p^2t+ims_1}\chi_+(t,\lambda)-
(r^2q_0)^{-\frac{1}{4}}e^{i\lambda p^2t+ims_2}\chi_-(t,\lambda),&m<0\ea\r. =\nn\\
\ll\{\ba{lr}
p^{-1}|t-t_c|^{-\fr}e^{-i\lambda p^2t-i\lambda p s_2\cos(\pi-\alpha)}\chi_+(t,\lambda)-&\\
p^{-1}|t-t_c|^{-\fr}e^{i\lambda p^2t+i\lambda ps_1\cos\alpha}\chi_-(t,\lambda),&m>0\\
p^{-1}|t-t_c|^{-\fr}e^{-i\lambda p^2t-i\lambda ps_1\cos\alpha}\chi_+(t,\lambda)-&\\
p^{-1}|t-t_c|^{-\fr}e^{i\lambda p^2t+i\lambda p s_2\cos(\pi-\alpha)}\chi_-(t,\lambda),&m<0\ea\r.\;\;\;\;\;\;\; =\nn\\
\ll\{\ba{lr}
i\Psi^-(t,s_2)-i\Psi^+(t,s_1),&m>0\\
i\Psi^-(t,s_1)-i\Psi^+(t,s_2),&m<0\ea\r.
\label{33c}
\ee
where $\Psi^\pm(t,s_j),\;j=1,2$, have the form \mref{17} but are exact. We have also assumed that
$(t-t_c)^{-\fr}=-i|t-t_c|^{-\fr}$ for $t<t_c$ (see below).

The last result shows that, depending on $m$, the bundles on which the fundamental solutions $\psi_\pm(r)$
are defined are chosen by the latter accordingly to their signatures.

Let us now consider the analytical continuation of the semiclassical limit $\Psi^{as}(r,\Phi)$ of the solution
$\Psi(r,\Phi)$ from the point $A$ to its point $B$ as it is defined by the rule \mref{33aa} or back, i.e. from $B$ to $A$
correspondingly to the rule \mref{33ba}, see Fig.2.

Since $t=t_c-\frac{1}{p}\sqrt{(r-r_+)(r+r_+)}$ then, if $r$ rounds the turning point $C$ by $2\pi$ clockwise starting
from the segment $AC$ (see Fig.13),
$t$ rounds $t_c$ on the $t$-plane also clockwise by $\pi$. But $t$ rounds $t_c$ by $\pi$ anticlockwise if $r$ does
the previous motion in the opposite direction. In both the cases we are found at the point $B$ with $t>t_c$. However in
the first case $(t-t_c)^{-\fr}=|t-t_c|^{-\fr}$ while in the second $(t-t_c)^{-\fr}=-|t-t_c|^{-\fr}$.

$\Psi^{as}(r,\Phi)$ continued clockwise on the $r$-Riemann surface is given by:
\be
\Psi_{clock}^{as}(r,\Phi)=\frac{\psi_+^{as}(r)-\psi_-^{as,cont}(r)}{r^\fr}e^{im\Phi}=
\frac{1}{r^\fr}\ll(1-ie^{-2i\lambda\int_1^{r_+}q^\fr dr'}\frac{\chi_{{\bar 1}\to 0}^{as}}{\chi_{1\to 0}^{as}}\r)
\psi_+^{as}(r)e^{im\Phi}\nn\\
r_+<r<1
\label{34b}
\ee
where $\psi_-^{as,cont}(r)$ has been substituted by its form which follows from \mref{30b}.

Continued however anticlockwise it is given by:
\be
\Psi_{anticlock}^{as}(r,\Phi)=\frac{\psi_+^{as,cont}(r)-\psi_-^{as}(r)}{r^\fr}e^{im\Phi}=
\frac{1}{r^\fr}\ll(1+ie^{+2i\lambda\int_1^{r_+}q^\fr dr'}\frac{\chi_{1\to 0}^{as}}{\chi_{{\bar 1}\to 0}^{as}}\r)
\psi_-^{as}(r)e^{im\Phi}\nn\\
r_+<r<1\;\;\;\;\;\;
\label{34c}
\ee
where $\psi_+^{as,cont}(r)$ has been substituted by its form given by \mref{30a}.

Note that in both the above continuations we have not taken into account the quantization conditions \mref{33aa}
and \mref{33ba}.

However if the energy is to be quantized we have to have:
\be
\Psi_{clock}^{as}(r,\Phi)=\Psi_{anticlock}^{as}(r,\Phi)\equiv 0
\label{34a}
\ee

Performing now in \mref{34b} and \mref{34c} the calculations similar to the ones above but with the variables
$t,\;t>t_c,$
we get from \mref{33c} and \mref{34b}, see Fig.7:
\be
\Psi_{clock}^{as}(r,\Phi)=\ll\{\ba{lr}
-i\Psi^{+,cont}(t,s_1)+i\Psi^-(2t_c-t,s_2),&m>0\\
-i\Psi^{+,cont}(t,s_2)+i\Psi^-(2t_c-t,s_1)),&m<0\ea\r.=\nn\\
\ll\{\ba{lr}-ip^{-1}|t-t_c|^{-\fr}\frac{\chi_{{\bar 1}\to 0}^{as}}{\chi_{1\to 0}^{as}}
e^{i\lambda p^2t+i\lambda ps_1\cos\alpha}\chi_+^{as}(t,\lambda)+&\\
p^{-1}|t-t_c|^{-\fr}e^{-i\lambda p^2(2t_c-t)-i\lambda ps_2\cos(\pi-\alpha)}\chi_+^{as}(2t_c-t,\lambda),&m>0\\
-ip^{-1}|t-t_c|^{-\fr}\frac{\chi_{{\bar 1}\to 0}^{as}}{\chi_{1\to 0}^{as}}
e^{i\lambda p^2t+i\lambda ps_2\cos(\pi-\alpha)}\chi_+^{as}(t,\lambda)+&\\
p^{-1}|t-t_c|^{-\fr}e^{-i\lambda p^2(2t_c-t)-i\lambda ps_1\cos\alpha}\chi_+^{as}(2t_c-t,\lambda),&m<0\ea\r.\nn\\
2t_c>t>t_c,\;\;\;\;\;s_2=s_1\pm2\alpha
\label{35}
\ee
where $\Psi^{+,cont}(t,s_j),\;t>t_c$, are the results of the continuations of
$\Psi^+(t,s_j),\;t<t_c,\;j=1,2$, by the caustic.

For $\Psi_{anticlock}^{as}(r,\Phi)$ we get instead:
\be
\Psi_{anticlock}^{as}(r,\Phi)=\ll\{\ba{lr}
-i\Psi^+(2t_c-t,s_1)+i\Psi^{-,cont}(t,s_2),&m>0\\
-i\Psi^+(2t_c-t,s_2)+i\Psi^{-,cont}(t,s_1)),&m<0\ea\r.=\nn\\
\ll\{\ba{lr}-p^{-1}|t-t_c|^{-\fr}e^{i\lambda p^2(2t_c-t)+i\lambda ps_1\cos\alpha}\chi_-^{as}(2t_c-t,\lambda)+&\\
-ip^{-1}|t-t_c|^{-\fr}\frac{\chi_{1\to 0}^{as}}{\chi_{{\bar 1}\to 0}^{as}}
e^{-i\lambda p^2t-i\lambda p s_2\cos(\pi-\alpha)}\chi_-^{as}(t,\lambda),&m>0\\
-p^{-1}|t-t_c|^{-\fr}e^{i\lambda p^2(2t_c-t)+i\lambda p s_2\cos(\pi-\alpha)}\chi_-^{as}(2t_c-t,\lambda)+&\\
-ip^{-1}|t-t_c|^{-\fr}\frac{\chi_{1\to 0}^{as}}{\chi_{{\bar 1}\to 0}^{as}}
e^{-i\lambda p^2t-i\lambda ps_1\cos\alpha}\chi_-^{as}(t,\lambda),&m<0\ea\r.\nn\\
2t_c>t>t_c,\;\;\;\;\;s_2=s_1\pm2\alpha
\label{35a}
\ee
where $\Psi^{-,cont}(t,s_j),\;t>t_c$, are the results of the continuations of
$\Psi^-(t,s_j),\;t<t_c,\;j=1,2$, by the caustic.

From the above considerations it follows clearly that $\Psi^+(t,s)$ continued through the caustic has to avoid the
focusing point at $t=t_c$ from above in the $t$-plane, moving clockwise while $\Psi^-(t,s)$ - from below, moving
anticlockwise. By such a continuation $\Psi^+(t,s)$ acquires the factor $i$ while $\Psi^-(t,s)$ - the factor $-i$.

Let us summarize the above results.
\begin{enumerate}
\item The SWF $\Psi^{as}(x,y)$ for the circular billiards originally defined closely to the boundary of the billiards
and
vanishing on the boundary can be written as the combinations \mref{33c} of the two SWF $\Psi^\pm(t,s)$ of the opposite
signatures having the forms given by \mref{17} and defined on the two skeletons ${\bf B}$ and ${\bf B}^A$.
\item  It is the SWF $\Psi^+(t,s)$ which is defined on the skeleton {\bf B} which momenta being
tangential to the caustic of the bundle give the anticlockwise orientation of the caustic for $m>0$ but on the skeleton
${\bf B}^A$ for $m<0$. The SWF $\Psi^-(t,s)$ is then defined on the second skeleton in the respective cases.
\item  $\Psi^{as}(x,y)$ can be continued across the caustic avoiding the real singular point at $t=t_c$ on the $t$-plane
from above for $\Psi^+(t,s)$ and from below - for $\Psi^-(t,s)$ and getting then the forms \mref{35} or \mref{35a}.
\item If $\Psi^{as}(x,y)$ is to be quantized then $\Psi^{as,cont}(x,y)$ has to vanish identically independently of the
way (clockwise or anticlokcwise) it was continued. In such a case we have to have:
\be
\Psi^{\pm,cont}(t,s)=\Psi^\mp(2t_c-t,s\pm 2\alpha),\;\;\;\;\;t>t_c
\label{37}
\ee
leading to for $t=2t_c$:
\be
\frac{\chi_{{\bar 1}\to 0}^{as}(E,\lambda)}{\chi_{1\to 0}^{as}(E,\lambda)}
\exp\ll[i\sum_{n\geq 0}E_{n+1}\lambda^{-n}\int_0^{2t_c}\frac{1}{1+\ll(\frac{q}{q_0}\r)^\fr}dt'\r]
e^{i\lambda p^22t_c-2i\lambda\alpha p\cos\alpha+i\fr\pi}=1
\label{38}
\ee
where we have taken into account that $r(2t_c-t)=r(t)$.
\item In the limit $\lambda\to\infty$ the condition \mref{38} decays into the following ones:
\be
e^{i\lambda pd(s,\alpha)-2i\lambda\alpha p\cos\alpha+i\fr\pi}=1\nn\\
\frac{\chi_{{\bar 1}\to 0}^{as}(E,\lambda)}{\chi_{1\to 0}^{as}(E,\lambda)}
\exp\ll[i\sum_{n\geq 0}E_{n+1}\lambda^{-n}\int_0^{2t_c}\frac{1}{1+\ll(\frac{q}{q_0}\r)^\fr}dt'\r]=1\nn\\
d(s,\alpha)=2pt_c=2\sin\alpha
\label{38b}
\ee

Of course the condition \mref{38} coincides with \mref{29}.
\item The two degenerate SWF's $\Psi^{as,\pm}(x,y)$ in each point inside the ring $\cos\alpha<r\leq 1$ can be given by:
\be
\Psi^{as,+}(x,y)=
-i\Psi^+(t,s_1)+i\Psi^{+,cont}(2t_c-t,s_2-2\alpha)=\nn\\i\Psi^-(t,s_2)-i\Psi^{-,cont}(2t_c-t,s_1+2\alpha),\;\;\;\;\;\;m>0\\
\Psi^{as,-}(x,y)=
-i\Psi^+(t,s_2)+i\Psi^{+,cont}(2t_c-t,s_1-2\alpha)=\nn\\-i\Psi^{-,cont}(2t_c-t,s_2+2\alpha)+i\Psi^-(t,s_1),\;\;\;\;\;\;m<0\nn\\
t<t_c
\label{38c}
\ee

\end{enumerate}

\section*{Appendix C}

\hspace{15pt}It is shown in this appendix that $\delta_k(u,l)$ from the formula \mref{18a} and $\delta(u_j,l_j)$ from the formula \mref{22a} are
$s$-independent. To this end consider Fig.1  on which a mapping $h_k(s;u,l)$ of the arc $A_k(u,l)$ into an arc $A_{k'}(u',l')$ is defined
by the the bundle $B_k(u,l)$. According to this mapping we have:
\be
{\bf r}_0(h_k(s;u,l))={\bf r}_0(s)+{\bf D}(s;u,l)\nn\\
{\bf r}_0(s)\in A_k(u,l),\;\;\;\;\;\;{\bf r}_0(h_k(s;u,l))\in A_{k'}(u',l')
\label{C1}
\ee
where ${\bf D}(s;u,l)$ is a vector linking the point ${\bf r}_0(s)$ with the point ${\bf r}_0(h_k(s;u,l))$ of the billiards boundary.

Differentiating the equation \mref{C1} with respect to $s$ we get:
\be
\frac{\p h_k(s;u,l)}{\p s}\cos\beta(h_k(s;u,l))=\cos\beta(s)-D(s;u,l)\frac{\p\gamma_k(s;u,l)}{\p s}\sin\gamma_k(s;u,l)+\nn\\
\cos\gamma_k(s;u,l)\frac{\p D(s;u,l)}{\p s}\nn\\
\frac{\p h_k(s;u,l)}{\p s}\sin\beta(h_k(s;u,l))=\sin\beta(s)+D(s;u,l)\frac{\p\gamma_k(s;u,l)}{\p s}\cos\gamma_k(s;u,l)+\nn\\
\sin\gamma_k(s;u,l)\frac{\p D(s;u,l)}{\p s}
\label{C2}
\ee
where $ D(s;u,l)$ is the length of ${\bf D}(s;u,l)$.

Making further the proper linear combinations of the last equations we have finally:
\be
\frac{\p h_k(s;u,l)}{\p s}\cos\alpha_{k'}(h_k(s;u,l);u',l')=\cos\alpha_k(s;u,l)+\frac{\p D(s;u,l)}{\p s}\nn\\
-\frac{\p h_k(s;u,l)}{\p s}\sin\alpha_{k'}(h_k(s;u,l);u',l')=\sin\alpha_k(s;u,l)-D(s;u,l)\frac{\p\gamma_k(s;u,l)}{\p s}
\label{C3}
\ee
where we have taken into account the following relation between the angles involved:
\be
\alpha_{k'}(h_k(s;u,l);u',l')+\alpha_k(s;u,l)=\beta(h_k(s;u,l))-\beta(s)+2\pi\nn\\
\gamma_k(s;u,l)=\beta(s)+\alpha_k(s;u,l)
\label{C4}
\ee
which follows from Fig.1.

The independence of $s$ of $\delta_k(u,l)$ and $\delta(u_j,l_j)$ follows now easily from the first of the relations \mref{C3}.

\section*{Appendix D}

\hspace{15pt}The billiard Laplacean expressed by the variables $d,s$ has the following form:
\be
\bigtriangleup(d,s)=\ll(\frac{\cos^2\alpha}{J^2}+1\r)\frac{\p}{\p d^2}+\frac{1}{J^2}\frac{\p^2}{\p s^2}-
\frac{2\cos\alpha}{J^2}\frac{\p^2}{\p s\p d}+\nn\\
\ll(\frac{\cos\alpha}{J^3}\ll((\alpha'+\gamma')\cos\alpha-\frac{\gamma''}{\gamma'}\sin\alpha\r)+
\frac{1}{J^2}\ll(\frac{\gamma''}{\gamma'}\cos\alpha+\alpha'\sin\alpha\r)-\frac{\gamma'}{J}\r)\frac{\p}{\p d}-\nn\\
\ll(\frac{1}{J^3}\ll((\alpha'+\gamma')\cos\alpha-\frac{\gamma''}{\gamma'}\sin\alpha\r)+
\frac{\gamma''}{\gamma'}\frac{1}{J^2}\r)\frac{\p}{\p s}\nn\\
J=\sin\alpha-\gamma'd
\label{1A1}
\ee
while for the corresponding operator $J^{\fr}\cdot\bigtriangleup\cdot J^{-\fr}$ we get:
\be
J^{\fr}\cdot\bigtriangleup\cdot J^{-\fr}=\frac{3}{4}J^{-2}(\nabla J)^2-J^{-1}\nabla J\cdot\nabla -\fr J^{-1}\bigtriangleup J+\bigtriangleup\nn\\
(\nabla J)^2=\frac{{\gamma'}^2}{J^2}\ll(d^2-2d\sin\alpha+1\r)+\frac{2\gamma'}{J^3}(\alpha'\cos\alpha-d\gamma'')\cos\alpha+
\frac{1}{J^2}(\alpha'\cos\alpha-d\gamma'')^2\nn\\
\nabla J\cdot\nabla=-\ll(\frac{\gamma'}{J^2}\ll(d^2-2d\sin\alpha+1\r)+\frac{1}{J^2}(\alpha'\cos\alpha-d\gamma'')\cos\alpha\r)\frac{\p}{\p d}+\nn\\
\ll(\frac{1}{J^2}(\alpha'\cos\alpha-d\gamma'')+\frac{\gamma'\cos\alpha}{J^2}\r)\frac{\p}{\p s}\nn\\
\bigtriangleup J=\frac{1}{J^2}\frac{\p^2}{\p s^2}(\sin\alpha-\gamma'd)+\frac{2\gamma''\cos\alpha}{J^2}-\nn\\
\frac{\gamma'}{J^3}\ll((\alpha'\cos\alpha-d\gamma'')\cos\alpha-(\beta'\sin\alpha-d{\gamma'}^2)(\sin\alpha-\gamma'd)\r)-
\frac{\gamma'\cos\alpha}{J^3}(\alpha'\cos\alpha-d\gamma'')
\label{1A2}
\ee

For the circular billiards the corresponding Laplacean takes the forms:
\be
\bigtriangleup(d,s)=\frac{1}{J^2}\ll(d^2-2d\sin\alpha+1\r)\frac{\p^2}{\p d^2}+\frac{1}{J^2}\frac{\p^2}{\p s^2}-
\frac{2\cos\alpha}{J^2}\frac{\p^2}{\p s\p d}-
\frac{1}{J}\frac{\p}{\p d}-
\frac{\cos\alpha}{J^3}\frac{\p}{\p s}\nn\\
J=\sin\alpha-d
\label{1A3}
\ee
while the corresponding operator $J^{\fr}\cdot\bigtriangleup\cdot J^{-\fr}$ the form:
\be
J^{\fr}\cdot\bigtriangleup\cdot J^{-\fr}(d,s)=\frac{3}{4J^4}\ll(d^2-2d\sin\alpha+1\r)+\frac{1}{J^3}\ll(d^2-2d\sin\alpha+1\r)\frac{\p}{\p d}-
\frac{\cos\alpha}{J^3}\frac{\p}{\p s}+\nn\\
\frac{1}{2J^2}+\bigtriangleup(d,s)
\label{1A4}
\ee

For the rectangular billiards the corresponding Laplacean has the following two similar forms depending on the rectangular sides:
\be
\bigtriangleup(d,s)=\frac{1}{\sin^2\alpha}\ll(\frac{\p^2}{\p d^2}+\frac{\p^2}{\p s^2}\r)\mp\frac{2\cos\alpha}{\sin^2\alpha}\frac{\p^2}{\p s\p d}\nn\\
\bigtriangleup(d,s)=\frac{1}{\cos^2\alpha}\ll(\frac{\p^2}{\p d^2}+\frac{\p^2}{\p s^2}\r)\mp\frac{2\sin\alpha}{\cos^2\alpha}\frac{\p^2}{\p s\p d}
\label{1A5}
\ee
and of course $J^{\fr}\cdot\bigtriangleup\cdot J^{-\fr}\equiv\bigtriangleup$ for this case of the billiards.

\end{document}